\documentclass[twocolumn]{aastex63}
\usepackage[T1]{fontenc}
\usepackage{verbatim}
\usepackage{xcolor}
\usepackage{graphicx}
\usepackage{natbib}
\usepackage{hyperref}
\usepackage{bm}
\usepackage{amsmath}
\usepackage{amsfonts}
\usepackage{amssymb}
\usepackage{fontawesome}

\usepackage{booktabs}
\usepackage{multirow}
\usepackage[para]{threeparttable}

\usepackage{soul}
\newcommand{\refeq}[1]{Eq.~(\ref{eq:#1})}          
\newcommand{\refeqs}[2]{Eqs.~(\ref{eq:#1})--(\ref{eq:#2})}

\newcommand{\reffig}[1]{Fig.~\ref{fig:#1}}

\newcommand{\refsec}[1]{Sec.~\ref{sec:#1}}

\newcommand{\be}{\begin{equation}}
\newcommand{\ee}{\end{equation}}
\newcommand{\bea}{\begin{eqnarray}}
\newcommand{\eea}{\end{eqnarray}}
\newcommand{\vs}{\nonumber\\} 
\def\ba#1\ea{\begin{align}#1\end{align}}

\renewcommand{\[}{\left[}
\renewcommand{\]}{\right]}
\renewcommand{\(}{\left(}
\renewcommand{\)}{\right)}

\renewcommand{\ln}{\operatorname{ln}}

\def\hto{{ {\rm H}_{\rm D,2}}}
\def\hd{{ {\rm H}_{\rm D}}}
\def\hm{{ {\rm H}^-_{\rD}}}

\newcommand*{\rHt}{\rH_2}
\numberwithin{equation}{section}

\newcommand*{\inc}{\Delta}

\newcommand*{\ra}[1]{r_{\alpha}^{#1}}
\newcommand*{\rc}[1]{r_{m}^{#1}}
\newcommand*{\rx}[1]{r_{M}^{#1}}
\newcommand*{\racx}[3]{\ra{#1}\, \rc{#2}\, \rx{#3}}
\newcommand*{\rac}[2]{\ra{#1}\, \rc{#2}}

\newcommand*{\rde}{r_{\inc E}}

\newcommand*{\eh}{E_{\rm H}}

\newcommand*{\KE}{\text{K.E.}}

\newcommand*{\rH}{{\rm H}}
\newcommand*{\rD}{{\rm D}}

\newcommand*{\dkns}{\textsc{DarkKROME}}

\def\htp{{ \rH_{\rD, 2}^+}}

\def\ba#1\ea{\begin{align}#1\end{align}}

\def\({\left(}
\def\){\right)}

\def\[{\left[}
\def\]{\right]}

\def\re{{\rm e}}
\def\rp{{\rm p}}
\def\rH{{\rm H}}

\begin{document}

\title{Molecular Chemistry for Dark Matter II: Recombination, Molecule Formation, and Halo Mass Function in Atomic Dark Matter}
\shorttitle{Atomic Dark Matter Evolution}

\author[0000-0002-8677-1038]{James Gurian}
\email{jhg5248@psu.edu}
\affiliation{Institute for Gravitation and the Cosmos, The Pennsylvania State University, University Park, PA 16802, USA}
\affiliation{Department of Astronomy and Astrophysics, The Pennsylvania State University, University Park, PA, 16802, USA}

\author[0000-0002-8434-979X]{Donghui Jeong}
\email{djeong@psu.edu}
\affiliation{Institute for Gravitation and the Cosmos, The Pennsylvania State University, University Park, PA 16802, USA}
\affiliation{Department of Astronomy and Astrophysics, The Pennsylvania State University, University Park, PA, 16802, USA}
\affiliation{School of Physics, Korea Institute for Advanced Study (KIAS), 85 Hoegiro, Dongdaemun-gu, Seoul, 02455, Republic of Korea}

\author[0000-0002-0378-5195]{Michael Ryan}
\email{mzr55@psu.edu}
\affiliation{Institute for Gravitation and the Cosmos, The Pennsylvania State University, University Park, PA 16802, USA}
\affiliation{Department of Physics, The Pennsylvania State University, University Park, PA, 16802, USA}

\author[0000-0002-6498-6812]{Sarah Shandera}
\email{ses47@psu.edu}
\affiliation{Institute for Gravitation and the Cosmos, The Pennsylvania State University, University Park, PA 16802, USA}
\affiliation{Department of Physics, The Pennsylvania State University, University Park, PA, 16802, USA}

\date{\today}

\begin{abstract}
Dissipative dark matter predicts rich observable phenomena that can be tested with future large-scale structure surveys. As a specific example, we study atomic dark matter, consisting of a heavy particle and a light particle charged under a dark electromagnetism. In particular, we calculate the cosmological evolution of atomic dark matter focusing on dark recombination and dark-molecule formation. We have obtained the relevant interaction-rate coefficients by re-scaling the rates for normal hydrogen, and evolved the abundances for ionized, atomic, and molecular states using a modified version of {\sf Recfast++} (which we have released publicly \href{https://github.com/jamesgurian/RecfastJulia}{\faGithub}\footnote{\url{https://github.com/jamesgurian/RecfastJulia}}). We also provide an analytical approximation for the final abundances. We then calculate the effects of the atomic dark matter on the linear power spectrum, which enter through a dark-photon diffusion and dark acoustic oscillations. At the formation time, the atomic dark matter model suppresses halo abundances on scales smaller than the diffusion scale, just like the warm dark matter models suppress the abundance below the free-streaming scale. The subsequent evolution with radiative cooling, however, will alter the halo mass function further.
\end{abstract}


\section{Introduction}
About 84\% of the matter in the universe is dark, revealing itself only through its gravitational charge, which is its mass. Various large-scale structure observation campaigns \citep{Bennett_2013, Planck_2020, Alam_2021} show that the observed spectrum of large-scale structure is consistent with the cold dark matter (CDM) model, and detailed study of the Bullet Cluster collision shows that the dark matter self-interaction is much weaker than gas particles \citep{markevitch_direct_2004}. There are small-scale phenomena such as cored density-profiles of dwarf galaxies \citep{Ostriker_2003, de_Blok_2010}, the missing satellite problem \citep{Diemand_2008, Springel_2008, bullock2010notes} and the too-big-to-fail problem \citep{Boylan_Kolchin_2011, Boylan_Kolchin_2012} that might suggest a deviation from prevailing CDM model, such as self-interacting dark matter \citep{Spergel_2000} or warm dark matter with non-negligible velocity dispersion \citep{Bode_2001, Sommer_Larsen_2001, Hogan_2001}. None of them, however, have conclusive evidence, and more accurate accounting for baryonic physics can also yield some of these phenomena (see \citet{Brooks_2019} for a comprehensive review).

On the other hand, despite decades of searching through direct detection \citep{Agnese_2018,cdmscollaboration2013silicon, Akerib_2017}, indirect detection \citep{Albert_2017, Bergstr_m_2013} and pair creation  \citep{Boveia_2018, Mitsou_2015}, there is no definitive evidence that dark matter interacts with Standard Model particles. It may be that the dark matter is for all practical purposes completely secluded from the Standard Model. Although this prospect would doom many current detection programs, it still leaves open the possibility of a rich dark sector consisting of one or more new particles interacting via new forces. Because such a model predicts distinctive dark-matter halo structure, it is possible to constrain the dark matter self interactions using only the gravitational interaction which the dark matter must possess \citep{Buckley_2018}. 

Here, we consider the particular possibility often referred to as ``atomic dark matter'' \citep{goldberg_new_1986,Ackerman2009,Feng2009,kaplan_atomic_2010, kaplan_dark_2011,cyr-racine_cosmology_2013,cyr-racine/etal:2014,Fan2013,Cline2014,boddy_hidden_2016,2015PhRvD..91b3512F,2016JCAP...07..013F,2015JCAP...09..057R,Agrawal2017,Ghalsasi:2017jna}. In this model, the dark matter consists of a heavy and a light fermion charged under a ``dark electromagnetism''. This dark matter can form hydrogen-like bound states including molecules. From one perspective, this can be seen as a toy model which captures some of the essential physics of a more complexly interacting dark sector. This model can satisfy constraints on the dark matter self-interaction cross section from, for example, colliding galaxy clusters and dwarf galaxy density profiles \citep{agrawal_make_2017}. In fact, dark matter models which introduce a mechanism to suppress the linear matter power spectrum on small scales (including warm dark matter and atomic dark matter) may alleviate challenges facing the $\Lambda$CDM model on small scales \citep{bullock_small-scale_2017}.

The parameters of the atomic-dark-matter model are the mass of the heavy particle $M$, that of the light particle $m$, the coupling constant $\alpha$ and the ratio of the dark-photon temperature to the cosmic microwave background (CMB) temperature, $\xi = T_{\gamma,\rD}/T_\gamma$. Observations of Big-Bang Nucleosynthesis (BBN) and the CMB temperature anisotropies constrain the energy density in relativistic species other than the three known neutrino generations, parameterized as the change in the effective number of neutrino species $\Delta N_{\rm eff}$. The most recent such constraint is $\Delta N_{\rm eff}\lesssim0.2$ \citep{Planck2015}, which restricts $\xi \lesssim 0.4$. This upper limit is easy to accommodate since the dark-photon temperature need not be the same as the CMB temperature. In some regions of the parameter space, the atomic dark matter model allows formation of a double disk \citep{Fan/etal:2013}, but this high dark-photon temperature region is strongly constrained by the lack of dark acoustic oscillations (DAO) in the observed galaxy power spectrum \citep{cyr-racine/etal:2014}. 

One particularly interesting aspect of atomic dark matter is that the dark matter is dissipative: atomic dark matter halos can radiate away kinetic energy. This means that these halos can undergo collapse, fragmentation, and can eventually form dark compact objects. This raises the prospect of probing an atomic dark matter model through gravitational wave events from their compact-binary coalescence \citep{shandera_gravitational_2018}. Crucially, the Chandrasekhar mass $M_{\rm Ch}\propto1/M^2$ only depends on the mass of the dark `proton', and it can be much smaller than the nominal value $M_{\rm Ch}\simeq1.4M_\odot$ in the Standard Model. That is, the mass spectrum of merger events, particularly for the sub-solar mass region, can carry information about the physics of even a fully secluded dark sector. \citet{shandera_gravitational_2018} has showed that a range of model parameters can produce merger rates accessible to the current generation of advanced-LIGO detectors. More recently, \citet{singh_gravitational-wave_2020} has derived constraints on the dark Chandrasekhar mass and showed that $M>0.95\,{\rm GeV}$ with 99.9\% C.L., if the unusual, so called mass-gap, gravitational wave event GW190425 \citep{Abbott_2020} is interpreted as the merger of an atomic-dark-matter black hole binary. 

So far, both \citet{shandera_gravitational_2018} and \citet{singh_gravitational-wave_2020} have relied on estimates of the dark black hole binary parameters obtained by re-scaling the Population III star binary literature \citep{Abel_2001, Bromm_2004, Hartwig_2016}. To obtain more realistic constraints, we must tackle directly the complex physics of the formation and evolution of dark-matter structures in the atomic-dark-matter model. As a stepping stone toward that complete study, in this work we consider the cosmological evolution of atomic dark matter. \citet{cyr-racine_cosmology_2013} have previously studied the recombination of dark hydrogen, which we extend by including the formation of molecular states. In the Standard Model, molecular hydrogen is the dominant coolant for pristine (zero metallicity) minihalos with virial temperature below $10^4\,{\rm K}$. So, studying the molecular abundance evolution must be a crucial step in determining the dark-halo properties within the atomic-dark-matter paradigm.  

Another essential ingredient for these studies is the modification of the linear matter power spectrum introduced by the atomic dark matter. Specifically, atomic dark matter produces dark acoustic oscillations (DAO) due to the propagation of pressure waves in the tightly coupled dark-matter/dark-photon fluid, as well as dark diffusion damping due to the non-zero mean free path of the dark photon. Both effects lead to a suppression of power on small scales, as discussed in detail in \citet{cyr-racine_cosmology_2013}. While \citet{cyr-racine_cosmology_2013} studied a relatively high value for the temperature ratio, most extensively $\xi =0.37$, we expand the study by varying $\xi$ in particular to a small value, $\xi < 0.1$. We have also added the molecular processes. We have shown that although molecular processes do change the evolution of the ionization fraction, as long as dark molecular hydrogen is rare relative to dark atoms, the effect is too small to alter the atomic dark matter effects on the linear matter power spectrum. 

We find that varying $\xi$ impacts all effects studied. At a given dark photon temperature, this parameter alters the redshift and hence the dark matter number density at a given temperature, which controls freeze-out and molecule formation. Similarly, the diffusion and DAO scales depend on the horizon size at decoupling, which depends on $\xi$ through the decoupling redshift.

We finally investigate the effects of atomic dark matter on the halo mass function in the presence of a cutoff in the linear power spectrum. This approach has been validated by simulations of warm dark matter \citep{Schneider_2013, Angulo_2013} and is an appropriate ansatz for the formation of atomic-dark-matter halos at high redshift. 

This paper is the second in a series of three. In Paper I, \citet{darkchem}, we defined a re-scaling procedure to compute the molecular chemistry for atomic dark matter from known Standard Model results. Paper III, \citet{darkkrome}, describes \dkns, a modification of the chemistry package KROME \citep{Grassi2014} to include the chemistry of atomic dark matter. In Paper III we use \dkns$\,$ and the cosmological abundances derived in this paper to explore one-zone collapse scenarios for a range of atomic dark matter parameters.

This paper is organized as follows. In \refsec{DarkRecombination}, we present the numerical implementation of the cosmological evolution of atomic dark matter states using {\sf Recfast++} modified by including the relevant molecular processes (\refsec{implement}), the computational method for event rates of these reactions (\refsec{rescaling}), and the result and an analytic approximation (\refsec{results}). In \refsec{consequences}, we study the effects of the addition of atomic dark matter and molecules on the formation of large-scale structure by calculating the dark diffusion scale, the DAO scale (\refsec{linearpk}), and finally estimating the halo mass function at its formation time (\refsec{darkhalos}). We conclude \refsec{conclusion} by pointing out the path toward the complete study of the large-scale structure with atomic dark matter.

Throughout, we shall attach the subscript $\rD$ to the Standard Model symbols to refer to the corresponding dark-matter symbols. For example, $\re_\rD+\rp_\rD\to \rH_\rD+\gamma_\rD$ refers to the dark-recombination reaction, the dark-matter analogue of the Standard Model reaction: 
$\re+\rp\to \rH+\gamma$.

\section{Dark Recombination and Molecule Formation}
\label{sec:DarkRecombination}
\subsection{The rate equations and implementation}
\label{sec:implement}
We begin by briefly summarizing the apparatus necessary to solve for the
chemical evolution of atomic-dark-matter states, including the recombination and molecule formation, in an expanding universe. To do so, we have implemented the core\footnote{That is, we have not implemented extra heating sources due to dark-matter annihilation \citep{Chen/Kamionkowski:2004,Padmanabhan/Finkbeiner:2005} or magnetic fields \citep{Sethi/Subramanian:2005,Kunze/Komatsu:2014}.} of  {\sf Recfast++} \citep{Seager_1999, chluba_recombinations_2010} in Julia \citep{bezanson2012julia}, using the
{\sf DifferentialEquations} library \citep{DifferentialEquations.jl-2017}.
We present a few details necessary for applying this code, which computes the Standard Model recombination, to the atomic-dark-matter scenario in Appendix \ref{sec:app}. 

The {\sf Recfast++} code uses Peeble's effective three-level recombination \citep{Peebles68} approximation, where the recombination is treated entirely by tracking the non-equilibrium population in the lowest three ($1s$, $2s$ and $2p$) atomic states and the continuum. High optical depth in the early universe to Lyman-limit photons prevents the direct recombination to the ground (1s) state (case-B recombination), so the two $n=2$ states ($2s$ and $2p$) are populated as meta-stable
states. The $n=2$ states are populated with the rate given by case-B recombination coefficient $\alpha_{H_\rD}$ that includes recombination to higher levels followed by a cascade down to $n=2$. They leave the $n=2$ state by one of three processes: (A) the {\it forbidden} two photon decay ($2s\to 1s$), (B) redshifting of Lyman-$\alpha$ photons out of the resonance band, and (C) photoionization. Exploiting the meta-stability (chemical equilibrium) of the $n=2$ state, $dx_2/dt =
0$, it is possible to solve for the population of that state and then for the net change in the ionization fraction. This leads to the following equation \citep{Seager_1999}, whose right-hand-side terms (in square brackets) represent (i) a change of variables from $t$ to $z$ (time to redshift), (ii) a branching ratio between the ground-state recombinations (two photon decay and redshifted Lyman-$\alpha$) and all exits from the $n=2$ state, including photoionization, and (iii) the net Case-B recombination rate:
\begin{widetext}
\begin{equation}
{\frac{dx_{ \rp_\rD}}{dz}}
    = \left[-\frac{1}{H(z)(1+z)}\right]
    \left[
        \frac{\Gamma_{\alpha,\rD}
    +  A_{ 2\gamma,\rD} }{\Gamma_{\alpha,\rD} 
    + A_{2\gamma,\rD} + \beta_{ \rH_\rD}} \right]
   \left[ \beta_{ \rH_\rD} x_{\rH_{\rD}\rm I}{ e}^{-h\nu_{{\rm Ly\alpha},D}/k_B T_{ g,\rD}} -x_{ \re_\rD}x_{ \rp_\rD} n_{ \rH_\rD} \alpha_{ \rH_\rD} 
   \right],
\label{eq:newstandard_xe}
\end{equation}
\end{widetext}
with the quantities appearing in the equation defined as follows: $k_B$ is the Boltzmann constant, $h$ is the Planck's constant, $T_{\rm g,D}$ is the dark-matter gas temperature and $T_{\gamma,D}$ is the dark-photon temperature, $x_{ \rp_\rD} = n_{ \rp_\rD}/n_{ \rH_\rD}$ is the free proton fraction, $x_{ \re_\rD}$ the free electron fraction, $x_{\rH_\rD \rm I}$ is the neutral hydrogen fraction. Note that $x_{\re_\rD}$ and $x_{\rp_\rD}$ differ because we also include the complex states $\rH_\rD^-$,  $\rH_{\rD, 2}^+$, $\rH_{\rD,3}^+$ in
the chemical network. The free electron fraction is calculated as $x_{\re_\rD} \equiv x_{\rp_\rD} - x_{\rH^-_\rD} + x_{\rH_{\rD,2}^+} + x_{\rH_{\rD, 3}^+}$.
We also have $\alpha_{ \rH_\rD}$ the case-B recombination rate,  
$\Gamma_{\alpha,\rD}\equiv 8\pi H(z)/\left[n_{\rH_{\rm D}\rm I}\lambda_{{\rm Ly\alpha},\rD}^3\right]$
the rate at which redshifted Lyman-$\alpha_D$ photons escape the resonance band, $\nu_{\rm Ly\alpha_\rD}$, the Lyman-$\alpha_\rD$ frequency,
$\beta_{\rH_\rD}$ the photoionization coefficient, and $A_{2\gamma,\rD}$ the 2-photon decay rate.

The recombination coefficient is given by the fitting formula of \citet{Pequignot_1991}. This formula includes an empirical 1.14 ``fudge factor'' \citep{Seager_1999} which accounts for the neglected higher ($n>2$) level transitions. The fudge factor takes into account all electric dipole transitions from higher $(n,\ell)$ modes.
Because dark hydrogen energy and angular momentum structure is merely re-scaled (\refsec{rescaling}) from the Standard 
Model hydrogen, we expect the same fudge factor to be applied to the atomic dark matter case.
We calculate the photoionization coefficient $\beta_{ \rH_\rD}$ by using detailed balance: $\beta_{ \rH_\rD} = \alpha_\hd(2\pi m k_B T_{\rm g,\rD}/h^2)^{3/2}\exp(-(E_{\rH_\rD} - h\nu_{{\rm Ly\alpha},\rD})/kT_{\rm g,\rD})$, with $m$ the mass of the light fermion and $E_{\rH_\rD}$ the dark hydrogen binding energy. As recombination proceeds, the dark matter and
dark photon fall out of thermal equilibrium due to the inefficiency of 
the Compton-scattering-type energy transfer. This process can be described by solving the following equation for $T_{\rm g,\rD}$ simultaneously with the equation for $x_{ \rp_\rD}$ \citep{Seager_1999}: 
\begin{equation}
\label{eq:cooling}
 \frac{dT_{\rm g,D}}{dz} = \frac{8\sigma_{\rm T,D}a_{\rm D}
   T_{\gamma,\rD}^4}{3H(z)(1+z)mc}\,
      \frac{n_{\re_\rD}}{n_{\rm DM}}\,(T_{\rm g,D} - T_{\gamma,\rD})
    + \frac{2T_{\rm g,D}}{(1+z)}. 
\end{equation}
where 
$n_{\rm DM}\equiv n_{\rH_\rD}(x_{\re_\rD} + x_{\rp_\rD} + x_{\rm H I} + x_{\rH_{\rD, 2}}+ x_{\rH_{\rD, 2}^+}+x_{\rH_{\rD, 3}^+}+x_{\rH_{\rD}^-})$
is the total dark-matter particle number density, $\sigma_{\rm T,D}$ is the Thomson cross section and $a_\rD$ the radiation constant. We discuss the relationship between these constants in the dark-matter sector and the Standard-Model values in the following subsection (\refsec{rescaling}).

We solve the ionization fraction and matter temperature equations [\refeqs{newstandard_xe}{cooling}] simultaneously with the equations for the evolution of $\rH_D^-$, $\rH_{\rD, 2}^+$, $\rH_{\rD, 3}^+$, and $\rH_{\rD, 2}$. These reactions also require the addition of terms which we collectively denote by $\frac{d\Delta x_{\rm p_\rD}}{dz}$ to \refeq{newstandard_xe} to account for the effect of the molecular reactions on the free proton fraction. The reaction rates, $c_{\rH i}$, appearing below are as summarized in Table \ref{tab:reactions} ($i$ is the reaction number in the Table), and have been obtained by re-scaling standard model rates according to the procedure describe in \refsec{rescaling} and in \citet{darkchem}. The number density of neutral dark atoms is $n_{   \rH_\rD \rm I}\equiv n_{\rH_\rD}(1-x_{\rp_\rD}-2x_{\hto}-2 x_{\htp}-3 x_{\rH_{\rD, 3}^+}-x_{\hm  })$. The additional evolution equations are

\begin{widetext}

\begin{align}
&\frac{d\Delta x_{\rm p_\rD}}{dz} = - \frac{-c_{\rH7} n_{\rp_\rD} x_{\hm} -c_{\rm \rH8} n_{ \rH_DI} x_{ \rp_D} +  c_{\rm H9}x_{ \rH_{\rD, 2}^+} + c_{\rm H10}x_{ \rH_{\rm D,2}^+}n_{ \rH_\rD \rm I} - c_{\rm H15}x_{ \hto}n_{ \rp_\rD} }{H(z)(1+z)}\\%
&\frac{d x_{ \rH_\rD^-}}{dz} = -\frac{c_{\rm  \rH3}n_{   \rH_D \rm I}x_{  \re_\rD}  - c_{\rm \rH4}x_{ \rH^-_\rD} - c_{\rH5}n_{\rH_\rD \rm I}x_{\hm}- c_{\rH7} n_{\rp_\rD} x_{\hm}  }{H(z)(1+z)}\\%
&\frac{d x_{ \htp}}{dz} = -\frac{c_{\rm \rH8} n_{ \rH_\rD I} x_{ \rp_\rD}- c_{\rm H9}x_{ \rH_{\rD, 2}^+} -c_{\rm H10}x_{ \rH_{\rm D,2}^+}n_{ \rH_\rD \rm I} -c_{\rm H13} x_{\hto} n_{\htp}+ c_{\rm H15}n_{ \rp_\rD}x_{ \hto}   + c_{\rm 3B4}n_{ \rH_\rD \rm I}^2 x_{ \rp_\rD}}{H(z)(1+z)}\\%
&\frac{d x_{ \rH_{\rD, 3}^+}}{dz} = - \frac{c_{\rm H13} x_{\hto} n_{\htp} - c_{\rm H20} x_{ \rH_{\rD, 3}^+} n_{\rm e_D}}{H(z)(1+z)}\\%
&\frac{d x_{ \hto}}{dz} = -\frac{c_{\rm H5}n_{ \rH_\rD \rm I}x_{ \rH_\rD^-}   + c_{\rm H10}x_{ \htp}n_{ \rH_\rD \rm I}  -c_{\rm H15}x_{ \hto}n_{ \rp_\rD} +c_{\rm H20} x_{ \rH_{\rD, 3}}^+ }{H(z)(1+z)}\\%
 &\qquad \qquad -  \frac{- c_{\rH*}  n_{\rH_\rD \rm I}x_{\hto} + c_{\rm 3B1}n_{ \rH_\rD \rm I}^2 x_{ \rH_\rD \rm I}+ c_{\rm 3B2}n_{ \rH_\rD \rm I}^2 x_{ \hto}+ c_{\rm 3B3}n_{ \rH_\rD \rm I}^2 x_{ H_\rD}^+}{H(z)(1+z)}.
\end{align}
\end{widetext}

\begin{table*}
\begin{rotatetable*}
		\scriptsize
		\centering
		\begin{threeparttable}
		\begin{tabular}{l l c c c c c}
			\toprule 
			\multirow{2}{*}{\#} & \multirow{2}{*}{Reaction} & \multirow{2}{*}{Cross section source}  & \multirow{2}{*}{$\sigma$} & Re-scaling pre-factor & \multirow{2}{*}{b} &   \multirow{2}{*}{Additional notes} \\
			& & & & $g(\ra{},\rc{},\rx{})$ &  & \\
			\midrule
			1 & $p+e\rightarrow \rH + \gamma$ & \citet{Mo2010} & $\frac{\alpha^5}{\KE(\KE+\inc E)} $  & $\rac{2}{-2}$  & $-0.62,-1.15$ &  \tnote{a} \tnote{,} \tnote{b} \\ 
			2 & $\rH + \gamma\rightarrow p+e$ & \citet{Mo2010} & $\mu\, \alpha^5\frac{1}{(\KE+\inc E)^3}$ & $\rac{5}{}$ & 0.88, 0.35 & \tnote{c}\\
			3 & $\rH+e \rightarrow \rH^- + \gamma$   & \citet{deJong1972}  & $\frac{ \alpha}{\mu^2}\; \frac{\inc E^{1/2}\, \KE^{1/2}}{(\KE+\inc E)} $ & $\rac{2}{-2}$  & 0.928  & \tnote{a} \tnote{,} \tnote{d} \\ 
			4 & $\rH^- + \gamma \rightarrow \rH + e$ & \citet{Armstrong1963} & $ \frac{\alpha}{\mu}\frac{\inc E^{1/2} K^{3/2}}{(\KE+\inc E)^3}$ & $\rac{5}{}$  &  2.13 & \tnote{c} \tnote{,}  \tnote{d} \\
			5 & $\rH^- + \rH \rightarrow \rHt+ e$ & \citet{Browne1969} & $ \sqrt{\frac{\alpha \, a_0^3}{\KE}} $ & $\racx{-1}{-3/2}{-1/2}$  & 0 & \tnote{e} \tnote{,} \tnote{f} \tnote{,} \tnote{g} \\
			7 & $\rH^- + p \rightarrow 2 \rH$   & \citet{Bates1955}  & $\alpha\,a_0^2 \sqrt{\mu} \, \frac{\sqrt{\KE+\inc E} }{\KE \, \inc E}$ &  $\rac{-3}{-3}$ & $-\frac{1}{2}$ & \tnote{e} \tnote{,} \tnote{h} \\
			8 & $\rH + p \rightarrow \rHt^+ + \gamma$ & \citet{Stancil1993} & $\frac{({\KE} +\inc E)^3 \alpha^{4}}{\eh^{3}{\KE}^{3/2} M^{1/2}}$ & $\racx{2}{-1}{-1}$  & 1.8 & \tnote{i}\\
			9 & $\rHt^+ + \gamma \rightarrow \rH + p $  & \citet{Stancil1993} & $\left(\frac{\mu \, v}{h\, \nu}\right)^2\,\frac{({\KE} +\inc E)^3 \alpha^{4}}{\eh^{3}{\KE}^{3/2} M^{1/2}}$ & $\racx{5}{1/2}{1/2}$  & 1.59 & \tnote{c} \tnote{,} \tnote{j}\\
			10 & $\rHt^+ + \rH \rightarrow \rHt+p$ &\citet{galli_chemistry_1998}  & $\sqrt{\frac{\alpha \, a_0^3}{\KE}} $ & $\racx{-1}{-3/2}{-1/2}$  & 0 & \tnote{g}\\
			13 & $\rHt^+ + \rHt \rightarrow \rH_{3}^+ +\rH$ &---\textquotedbl--- & ---\textquotedbl--- & ---\textquotedbl---   & 0 & \tnote{g}\\

			15 & $\rHt + p \rightarrow \rHt^+ + \rH$ & ---\textquotedbl--- & ---\textquotedbl--- & ---\textquotedbl---    & 0 &  \tnote{g} \tnote{,} \tnote{j}\\
			20 & $ \rH_{3}^+ + e\rightarrow \rH +\rHt $ &\citet{Draine2011} & $\frac{\alpha \, a_0}{\KE}$ & $\racx{-1}{-2}{}$  & -0.65 & \tnote{k}\\
			* & $\rHt + \rH \rightarrow 3 \rH$  & Hard Sphere & $a_0^2$ & $\racx{-1}{-3/2}{-1/2}$ & 0 & \tnote{l} \\
			3B1 & $3\rH \rightarrow \rHt+\rH$  & Hard Sphere/Detailed Balance & $a_0^2\left[\frac{n_{\rHt}}{n^2_\rH}\right]_{\rm LTE}$ & $\racx{-4}{-4}{-1}$ & -1 & \tnote{j} \tnote{,} \tnote{m} \\
			3B2 & $\rHt+2\rH \rightarrow 2\rHt$  & ---\textquotedbl--- & ---\textquotedbl--- & ---\textquotedbl---  & -1 & \tnote{j} \tnote{,} \tnote{m} \\
			3B3 & $2\rH + \rH^+ \rightarrow \rHt+\rH^+$  &---\textquotedbl---& ---\textquotedbl--- &---\textquotedbl---  & -1& \tnote{j} \tnote{,} \tnote{m} \\
			3B4 & $2\rH + \rH^+ \rightarrow \rHt^+ + \rH$ & ---\textquotedbl---& ---\textquotedbl---& ---\textquotedbl---  & -1 & \tnote{j} \tnote{,} \tnote{m} \\
			\bottomrule
		\end{tabular}
\caption{Table of Reactions. Reactions are numbered according to \citet{galli_chemistry_1998} and the section numbers in parentheses indicate where details can be found in the text. Only the parametric dependence on key quantities are shown for the cross sections (no numerical factors). The reactions included in this table are those that were considered ``important" in \citet{galli_chemistry_1998} and additions discussed in the text.  To leading order the Standard Model rates can be approximated as $\gamma \propto g(\ra{},\rc{},\rx{}) \; (T/\rde)^b$ and values for $b$ are from Galli and Palla unless otherwise noted. All reactions listed have binding energy proportional to $\eh$, so here $\rde=\rac{2}{}$. For reactions 1-4 and 20, the reduced mass, $\mu$, is proportional to the dark electron mass, $m$. For the remaining reactions, $\mu$ is proportional to the dark proton mass, $M$.
\label{tab:reactions}}
		\begin{tablenotes}
			\item [a] Milne Relation
			\item [b] The $b$ values are an expansion of the Case B recombination coefficient in low and high temperature regimes (respectively), from \citet{Pequignot1991}. The re-scaling of the Case A coefficient is identical, although $b$ differs. 
			\item [c] Photoionization re-scaling
			\item [d] Effective Range approximation
			\item [e] These rates are uncertain by up to an order of magnitude \citet{Glover2008}
			\item [f] Rate is constant for $T\le300{\, \rm K}$, and very uncertain for $T>300{{\, \rm K}}$ \citet{galli_chemistry_1998}
			\item [g] Langevin Reaction
			\item [h] Landau-Zener method
			\item [i] $b$ value from \citet{Stancil1998}
			\item [j] General Detailed Balance
			\item [k] Coulomb focusing
			\item [l] The $b$ value is from \citet{Lepp1983} and does not account for density dependence.
			\item [m] The $b$ value is from \citet{Yoshida_2006}.
		\end{tablenotes}
		\end{threeparttable}
\end{rotatetable*}
\end{table*}
	
Most reactions in Table \ref{tab:reactions} are those of the minimal model in \cite{galli_chemistry_1998}, and the rates are those compiled in the same reference, except that we replace the more complicated temperature dependence of H7 and H8 in \cite{galli_chemistry_1998} with the simpler power laws taken from \cite{Bates1955} and \cite{Shapiro}, respectively.

\subsection{Re-scaling the reaction rates}
\label{sec:rescaling}
We have re-scaled the recombination rate with different values of $m_e$, $m_p$, and $\alpha$ as described in \cite{Hart_2017}:
\begin{align}
\label{eq:recfast_scaling}
\begin{split}
    &\sigma_{ \rm T,D} \propto r_\alpha^2 r_{m}^{-2} 
\qquad A_{2\gamma,\rD} \propto r_\alpha^8 r_m 
    \qquad \Gamma_{\alpha,\rD} \propto r_\alpha^{6} r_m^{3}
\\
    &
\alpha_{ \rH_\rD} \propto r_\alpha^2 r_m^{-2} 
\qquad \beta_{ \rH_\rD} \propto r_\alpha^5 r_m\,
\end{split}
\end{align} 
where $r_x$ denotes the ratio of a dark quantity to its standard model value, for example $r_M = M/m_p$. 
Here, the re-scalings of $\sigma_{T,D}$, $A_{2\gamma,D}$, and 
$\Gamma_{\alpha D}$ are immediate from the scaling of the hydrogen binding
energy, $E_{\rH} \propto r_m r_\alpha^2$. The scalings of $\alpha_{ \rH_\rD}$ and $\beta_{ \rH_\rD}$ are derived in \cite{darkchem} and \cite{Hart_2017}. 

Additionally, we take into account the temperature dependence of the
recombination and photoionization coefficients by re-scaling the temperature
to keep the temperature-to-hydrogen-binding-energy ratio invariant:
\be
T_\rD = \frac{T}{E_\rH} E_{\rH,\rD} = r_\alpha^{2}r_mT\,.
\ee
When the functional form is given in terms of the Standard-Model temperature,
we re-scale the dark-matter temperature to find the corresponding 
Standard-Model temperature by $T=r_\alpha^{-2}r_m^{-1}T_D$ for evaluating the rate coefficient. This re-scaled temperature is denoted $\tilde{T}_a$.
For example, we calculate the recombination coefficient by
\begin{equation}
\alpha_{\rH_\rD}(T_\rD) = r_\alpha^2 r_m^{-2} \alpha_\rH(r_\alpha^{-2}r_m^{-1}T_\rD) =r_\alpha^2 r_m^{-2} \alpha_\rH(\tilde{T}_a) \,.
    \label{eq:rescaledT}
\end{equation}

For the reactions involved in the formation of molecular states we adopt re-scaling procedure described in detail in \cite{darkchem}. The reaction rates in Table \ref{tab:reactions} are fits to either theoretical calculations or experimental data. The form emerges from integrating the cross section over a Maxwell-Boltzmann distribution (or, for radiative reactions, a Planck distribution) with a lower energy cutoff corresponding to the reaction threshold. As a result, the form of the re-scaled rate is given as 
\begin{equation}
    c_{\rm Hi,D} = {g(r_\alpha, r_m, r_M)}c_{\rm Hi}(\tilde{T}_a)
\end{equation}
where $g(r_\alpha, r_m,r_M)$ contains the entire parametric dependence of the cross section.

The reactions 1 to 15 in Table \ref{tab:reactions} are considered ``important" in \citet{galli_chemistry_1998}. One might worry that other reactions deemed unimportant in the Standard Model by \cite{galli_chemistry_1998} could also contribute for nonstandard parameters. This issue is discussed in more detail in \cite{darkchem}. Briefly, the volumetric rates $c_{\rm Hi}$ of the unimportant reactions may be small either due to a strong parametric suppression of the rate coefficient $c{\rm Hi}$ or by the small relative abundance of the reactants. 
Overcoming these large order-or-magnitude differences requires either an 
extreme-valued re-scaling factor $g(r_\alpha,r_m,r_M)$ outside of the range that
we consider here, or a similarly large modification of the relative number densities. Such a modification of the relative abundances can occur when $\xi$ is small, where recombination occurs in a strongly collisional environment. For this reason, we have included the reaction $*$ for the collisional destruction of $\rH_2$, as well as the three body reactions relevant to the formation of $\rH_2$ used in \cite{Yoshida_2006}. We also include the dominant $\rH_{\rD, 3}^+$ formation and destruction mechanisms, which are suppressed for the Standard Model only by the low fractional abundance of $\rH_2$. While reasonable, these modifications may be insufficient in some parameter regimes. In the Standard Model literature, the minimal reaction list of \cite{galli_chemistry_1998} has been largely validated by later, exhaustive studies (i.e. \cite{Glover2009}). Pending similar work for the dark sector, some caution is warranted when the relative abundances of the species are inverted by orders of magnitude compared to the Standard Model values.

\subsection{Limitations of Re-scaling}
As we have discussed in \cite{darkchem}, the re-scaling procedure has the following limitations. 

First, the theoretical calculations of the molecular cross 
sections in the Standard Model rely on a Born-Oppenheimer (stationary nuclei) 
approximation that fails when $m \approx M$. 

Second, some theoretical cross 
sections are computed under assumptions of thermal equilibrium between matter and radiation and/or chemical equilibrium between excited states. This assumption is not strictly justified either for the Standard Model or for atomic dark matter, but is adopted regardless for simplicity. For example, \cite{cyr-racine_cosmology_2013} has pointed out that weakly coupled dark matter can thermally decouple from the radiation prior to recombination, clearly breaking the assumption that each reaction is a function of
only one temperature. 
That work has also derived a recombination cross-section with stimulated recombination, which explicitly depends on the radiation temperature, and 
showed that the stimulated recombination yields less than 60\% change in the
final free electron fraction. This level of difference is small compared to the difference
due to the overall uncertainty in the reaction rates. 
The suggested rates in the literature for a single reaction 
can differ by an order of magnitude or more \citep{savin_rate_2004}. 

Finally, the accuracy of the re-scaling method is limited by the accuracy of the
Standard Model interaction rates.
In this work we do not attempt to improve upon the uncertainties in the Standard Model literature. Instead, we scrutinize the sensitivity of 
the final free electron fraction and molecular hydrogen abundance on the 
interaction rates.
In \refsec{results_analytic}, we show that the uncertainties in the important 
reaction rates propagate through the calculation in a simple manner 
and demonstrate that the uncertainties at worst linearly affect the final free 
electron fraction and molecular abundance (\reffig{uncert}). 

There are additionally three conditions which limit the range of parameters within which our results can be trusted. The first is that if recombination occurs in a sufficiently high density environment, generally quoted around $10^8\, \rm cm^{-3}$ in the Standard Model, three body reactions will become important \citep{Palla1983}. We can similarly find the density at which three-body reactions become important by equating the two-body 
interaction timescale 
\begin{equation}
    t \sim (n\sigma v)^{-1},
\end{equation}
with the three-body interaction timescale
\begin{equation}
    t \sim (n^2\sigma^{5/2} v)^{-1}. 
\end{equation}
Then, scaling the density at recombination as 
$n \sim 10^3 \xi^{-3} r_m^{3} r_\alpha^6  r_M^{-1}$ 
and $\sigma\propto a_0^2$ with the Bohr radius $a_0$, 
we find the following condition for three-body reactions to be unimportant 
at recombination time:
\begin{equation}
    \xi^{-3} r_M^{-1} r_\alpha^{3} < 10^5.
    \label{eq:tblimit}
\end{equation}
We have included a minimal set of three-body reactions, but as discussed above some caution is still warranted in the parameter regime where Eq.~(\ref{eq:tblimit}) is violated. 

A second limiting condition arises from the spectral distortions in the dark photon field induced by the recombination process. In the Standard Model, \cite{Hirata06} showed that the excess Lyman-$\alpha$ and two-photon process contributions to the radiation field suppress the formation of $\rH_2$ by a factor of $\sim5$ compared to naive calculations. This suppression is primarily due to ionization of $\rH^-$ by the redshifted spectral distortions, which dominate the radiation field at those frequencies. The size of this effect is thus controlled by the number of spectral distortion photons compared to the number of $\rH^-$ ions. In the Standard Model, roughly 5 photons are produced per hydrogen atom during the recombination process \citep{Chluba_2006}. As long as this number does not depend sensitively on the dark parameters, we do not expect any enhancement of this effect and neglect it for simplicity. Since the energy and angular momentum structure of the dark hydrogen are the same as ordinary hydrogen, we do not expect this number to change as long as the radiation field remains approximately black-body.

However, this is not guaranteed for all values of the parameters. The Standard Model baryon-to-photon ratio $\eta$ is order $10^{-10}$, which means that the photons produced by the recombination process do not substantially alter the radiation field, other than through the Lyman-$\alpha$ and two photon processes which we explicitly include. Manipulating $\xi$ and $M$ can produce a much larger value of $\eta_\rD$. To handle the recombination and molecule formation problems in this case would require a full non-equilibrium treatment of the spectral distortions, since the recombination photons will destroy the black-body nature of the radiation field. To determine when this may occur, we start from the fact that for the Standard Model the spectral distortions due to higher hydrogen transitions are
\be
\frac{\Delta I_\nu}{B_\nu} \lesssim 10^{-7},
\ee
with $\Delta I_\nu$ the intensity of the spectral distortion and $B_\nu$ the black-body intensity. We insist that the distortions remain less than order unity and insert the scaling of $\eta_D$ to obtain
\be
\xi^{-3}r_M^{-1} < \mathcal{O}(10^7).
\label{eq:sdlimit}
\ee

Finally, in the Standard Model, the recombination temperature is lower than the
binding energy of hydrogen because of the low cosmological number density. Since the gas temperature is set by the CMB, this density is parameterized by the baryon-to-photon ratio $\eta$.
Even the weak (logarithmic) dependence of the recombination redshift on
$\eta$ eventually pushes the recombination temperature outside the regime of
validity of some of our interaction rates. 
We omit the parameters for which the re-scaled temperature (see \refeq{rescaledT}) at the recombination redshift is greater than $10^4$ K. This is the highest temperature at which all of the reaction rate laws remain valid. We do not consider parameters where the re-scaled temperature at recombination is much less than in the Standard Model. 

When $r_\alpha \gtrsim 10^{-2/3}\simeq 0.2$, satisfying \refeq{tblimit}, the three body limit, automatically satisfies \refeq{sdlimit}, the spectral distortion condition. For smaller $r_\alpha$, the opposite is true. The third condition, from the temperature range of the reaction rates, seems never to be a problem when the first two conditions are met.
It is worth emphasizing that these are practical rather than theoretical limitations. In theory, there is no obstacle to the dark recombination occurring at arbitrarily high $\eta_\rD$. 

\subsection{Analytic estimation}
\label{sec:results_analytic}
We now develop some analytic approximations for the final abundance of molecular states to help us digest the numerical results in \refsec{results} and quantify the propagation of the uncertainties in reaction rates. 
Our approach is based on \cite{lepp_molecules_1984} (see also \cite{stiavelli}).

We begin with the two pathways for the formation of molecular hydrogen known from the Standard Model, the $\htp$ channel: 
\ba
&  \rH_\rD + \rH_\rD^+ \rightarrow \rH_{\rD, 2}^+ + \gamma\\ 
&  \rH_{\rD, 2}^+ + \rH_\rD \rightarrow \rH_{\rD, 2} + \rH^+\,,
\ea
and the $\rH^-$ channel:
\ba
&  \rH_\rD + \re_\rD \rightarrow \rH^-_\rD + \gamma\\
&  \rH_\rD^- + \rH_\rD \rightarrow \rH_{\rD, 2} + \re_\rD\,.
\ea
These reactions occur in chemical equilibrium:
\be
\frac{d x_{i}}{dt} = 0,
\label{eq:CEgen}
\ee
where $x_i$ refers to the fraction of $\rH_{\rD, 2}^+$ or $\rH_\rD^-$.
The equilibrium condition reduces the differential equation for the 
abundance $x_i$ to an algebraic equation. We further simplify this equation by considering only the dominant production and destruction channels for the species. Initially, the second reaction in each pathway is negligible compared to the radiative destruction of $\htp$ and $\hm$. For example, the chemical equilibrium equation for $\rH_\rD^-$ is 
\ba
\frac{dx_{\rH^-_\rD}}{dt} &= 0\nonumber \\ 
&= c_{\rH 3}x_{\re}n_{\rH_{\rm D}\rm I} - c_{\rH 4}x_{\rH_\rD^-} \,.
\label{eq:CE}
\ea
Here, we use the time variable $t$ to emphasize that this chemical equilibrium
holds on time scales much shorter than the background evolution time scale,
the Hubble time.

As the universe cools adiabatically, chemical equilibrium dictates that $x_i$ must increase so that the photo-destruction rate continues to balance the creation rate. 
At some critical redshift, however, ionizing photons can no longer efficiently
destroy the intermediate species such as $\rH_{2, \rD}^+$ and $\rH_\rD^-$. Then, chemical 
equilibrium exists not between creation of the intermediate and its 
photo-destruction but instead between creation of the intermediate and 
the formation of $\hto$. Since the reactions which produce $\hto$ are fast
compared to those which produce the intermediate, we assume that all of the
intermediate species present at this critical redshift are converted into 
$\hto$. 
No additional $\hto$ is produced via the given channel at lower redshifts.

That is, the $\rH_2$ produced from the channel $i$ is
\begin{equation}
    x_{\rH_{\rD, 2},i} = x_{i,\rm CE}(z_{\rm crit}),
    \label{eq:analytic}
\end{equation}
where $x_{i,CE}$ is the chemical equilibrium solution of \refeq{CEgen}.

The ${\rm H}_2$ production from each channel depends on 
the free electron abundance. We approximate the residual electron abundance of recombination 
as the free electron abundance available for the molecule formation.
The freeze-out time ($z_{\rm fo}$), when the recombination rate equals the 
Hubble expansion rate, sets the residual electron abundance as 
\be
c_{\rH1}(z_{\rm fo})n_{\rH_\rD} x_{\re}(z_{\rm fo})  = H(z_{\rm fo}).
\label{eq:fo}
\ee
 Here, we use $x_{\re}(z)$ from Saha equilibrium, that is, setting
the right hand side of \refeq{newstandard_xe} to zero, and neglecting the difference between $x_{e_\rD}$ and $x_{p_\rD}$:

\begin{align}
\frac{x_\re^2}{1-x_\re}
=&
\frac{\beta_{\rH_\rD}}{\alpha_{\rH_\rD} }\frac{1}{n_{\rH_\rD}}e^{-h\nu_{{\rm Ly\alpha},D}/k_BT_{\rm g,D}}
\nonumber
\\
=&
\frac{1}{n_{\rH_\rD}}\left(\frac{2\pi mk_BT}{h^2}\right)^{3/2}e^{-E_{\rH_\rD}/k_B T_{\rm g,D}}\,.
\label{eq:saha}
\end{align}

Some molecular hydrogen can be also produced during recombination, but in the
Standard Model the amount is very small and is usually neglected. For 
some dark parameters, however, this first burst of formation dominates the 
final molecular abundance. In particular, small $\xi$ triggers such a burst
because dark recombination happens earlier, so that the collisional 
$H_2$ formation processes are enhanced compared to the Standard-Model case. 
In this case, we estimate the formation of $\hto$ under the assumption of chemical equilibrium between the dominant creation and destruction reactions:
\ba
\frac{dx_{\hto}}{dt}  &= 0 \nonumber \\
&=c_{\rm H5} n_{ \rH_\rD \rm I} x_{ \rH_\rD^-} + c_{\rm H10} n_{ \rH_\rD \rm I}x_{ \rH_{\rD, 2}^+} \nonumber\\
&\quad- c_{\rm H15} n_{ \rH_\rD^+} x_{\hto} - c_{\rH*}  n_{\rH_\rD \rm I}x_{\hto},
\label{eq:recchannel}
\ea
where the abundances of $ \rH_{\rD, 2}^+$ and $ \rH^-_\rD$ are themselves evaluated using chemical equilibrium between their dominant creation and photodestruction processes and the expression is evaluated at $z_{\rm eff} = (z_{\rm rec} + z_{\rm dec})/2$. That is, midway between the redshift $z_{\rm rec}$ at which the Saha ionization fraction [\refeq{saha}] is $x_\re=0.5$ and the redshift of last scattering $z_{\rm dec}$, the peak of the visibility function $g(z)=(d\tau/dz)e^{-\tau(z)}$ with the
optical depth $\tau(z) = \int {\rm d}\ell n_e\sigma_T$.
This approximation relies on the fact that the $\hto$ production is still slow relative to the production and other (besides $\hto$ formation) destruction reactions for the intermediates. That is, the equilibrium point between creation and destruction of $\htp$ and $\hm$ is not strongly affected by this burst of $\hto$ formation.

For the efficiency of molecule formation during recombination, 
the particle number density $n_{\rH_\rD}$ at recombination turns out to be the single strongest controlling parameter. This is because, unlike atomic hydrogen, molecular hydrogen is mainly created and destroyed collisionally. 
In the Standard Model, by the time of recombination ($T\sim 0.25$ eV), it is
already energetically favorable to form molecules, and in fact we do see a
burst of molecule formation in all cases we study in \refsec{results}
[Fig.~\ref{fig:interesting_tracks}], as soon as neutral atoms are available.
Nonetheless, this formation channel is completely negligible for the
parameters close to the Standard Model.
In the Standard Model, this channel is inhibited by two factors: the radiative destruction of $\rH_{\rD, 2}^+$ and $\rH_\rD^-$ (they are fragile enough to be destroyed even at low temperatures), and the simple unlikeliness of the necessary encounters between species with only a small fractional abundance. 
Both problems are remedied by increasing the particle number density.
For a fixed $\rho_{\rm DM}$, the dark-particle-to-photon ratio scales as $\xi^3/r_M$. That is, for low $\xi$ recombination happens at very high redshift, and since $n \propto (1+z)^3$ the number density can be much higher compared to the Standard Model. The dependence on the other parameters is similar: the higher $m$ and $\alpha$ the higher the atomic binding energy and hence the redshift of recombination. As the density at recombination increases, we find that 
$x_{\rH_{\rD, 2}} \to 0.5$; that is, the dark-matter gas approaches a 
completely molecular state.

\subsection{Numerical results}
\label{sec:results}
\begin{figure*}
    \centering
    \includegraphics[width=\textwidth]{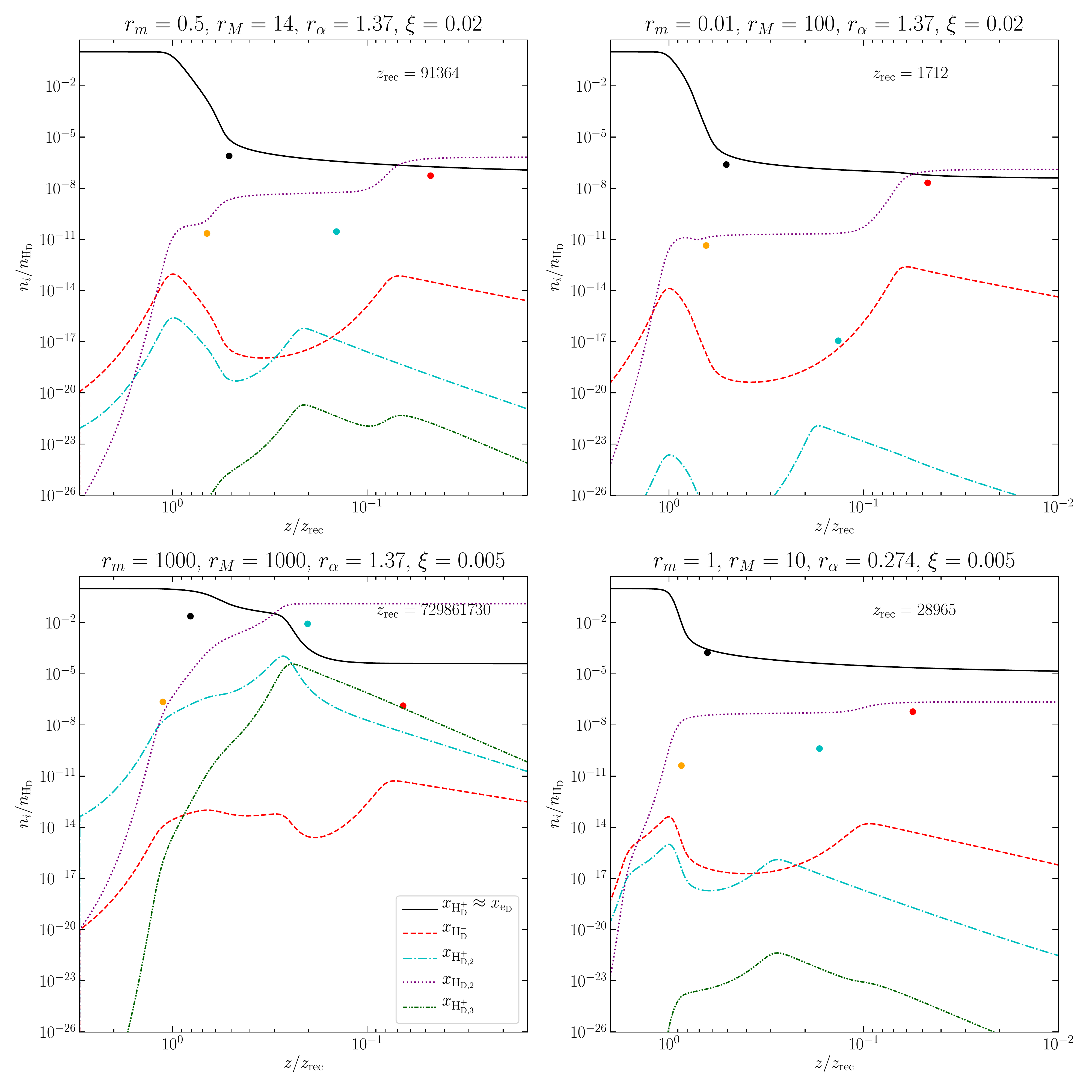}
    \caption{The evolution of the dark matter gas for several choices of parameters. The circles colored to match each line show our analytic approximations for the freeze out free proton fraction and $\rH_{\rD, 2}$ produced from each of the three colors (gold for the recombination channel, cyan for the $\rH_{\rD, 2}^+$ channel and red for the $\rH_\rD^-$ channel. If a previous channel has already created a large amount of molecular hydrogen these dots can be far from the $\rH_{\rD, 2}$ curve and still accurately depict the amount of $\rH_{\rD, 2}$ formed through that channel. }
    \label{fig:interesting_tracks}
\end{figure*}

In \reffig{interesting_tracks}, we show the cosmological evolution from the modified {\sf Recfast++} of 
the atomic dark matter states (ionized, atomic, and molecular) as a function of redshift, renormalized in the unit of $z_{\rm rec}$, the recombination redshift.
For each panel, we show the abundances of ${\rH}_\rD^+$ (black, solid line),
$\rH_{\rD, 2}^+$ (cyan, dot-dashed line),
$\rH_{\rD}^-$ (red, dashed line),
and $\rH_{\rD, 2}$ (gold, dotted line).
The redshift range we show here spans from the beginning of recombination to 
the end of molecule formation. 
The four panels in \reffig{interesting_tracks} show a range of parameters, which are representative of the range of possible evolution histories.

The dots correspond to our analytic estimates of the electron freeze out [\refeq{fo}] and each of the three channels for $\rH_{\rD, 2}$ production [\refeqs{analytic}{recchannel}]. Note that each dot indicates only the additional $\hto$ generated from a particular channel, so if this amount is much smaller than the preexisting $\hto$ abundance the dot will lie far from the gold curve. In that case, we have verified only that the estimate correctly predicts little enough $\hto$
production to negligibly alter the final abundance\textemdash not that this tiny fractional increase in $x_\hto$ agrees exactly with the numerical result.
As shown in \reffig{interesting_tracks}, the analytic approximation is generally accurate to within an order of magnitude for the final $\rH_{\rD, 2}$ fraction. 

The two top panels with $\xi = 0.02$ are qualitatively quite similar to Standard Model (e.g. \cite{galli_chemistry_1998, lepp_molecules_1984}) tracks. Note however that the recombination occurs at $z_{\rm rec}\sim 10^5$ in the top left panel, due to the reduced temperature. In the right panel, this is largely offset ($z_{\rm rec}\sim2000$) by the low mass $m$ reducing the hydrogen binding energy. By lowering $m$ or $\alpha$ (or increasing $\xi)$, recombination can be delayed to arbitrarily low redshift or pushed into the future. 
These parameter choices are less observationally relevant, 
since, unless the coupling $\alpha$ is extremely small, the ionized dark 
matter with the dark-electromagnetic interaction would fail to behave like CDM
on large scales.

The bottom two panels show more extreme parameters. The bottom left shows ``heavy'' dark matter with $\xi = 0.005$. Here, the recombination happens at $z\sim 10^9$ in a very high density environment. This precedes even Big-Bang Nucleosynthesis (BBN), but for completely secluded DM, BBN is not affected because the contribution to the total relativistic species from the dark matter is extremely small with $\xi=0.005$. In this case we find the asymptotic value of $x_{\hto} =0.13$. As we have discussed at the end of \refsec{results_analytic}, this reflects a general trend toward forming more molecular hydrogen when the formation occurs at high density. Here, the very large $\hto$ abundance allows the formation of $\rH_{\rD, 3}^+$ to compete with the formation of $\hto$ once the $\htp$ photo-dissociation becomes ineffective. The $\rH_{\rD, 3}^+$ is in turn efficiently converted into neutral hydrogen and $\hto$ by dissociative recombination. This leads to a net reduction in $x_{\rH_\rD^{+}}$.

Finally, the bottom right panel shows $r_\alpha = 0.274$ and $\xi= 0.005$, and $r_M = 10$. Here we see that even for small values of $\xi$ we can end up with the same qualitative story as in the Standard Model provided that the other parameters are adjusted to control the number density at recombination. Note however that the weak coupling leads to a relatively high freeze-out ionization fraction and relatively low $\hto$ fraction.

\begin{figure*}
    \centering
    \includegraphics[width=\textwidth]{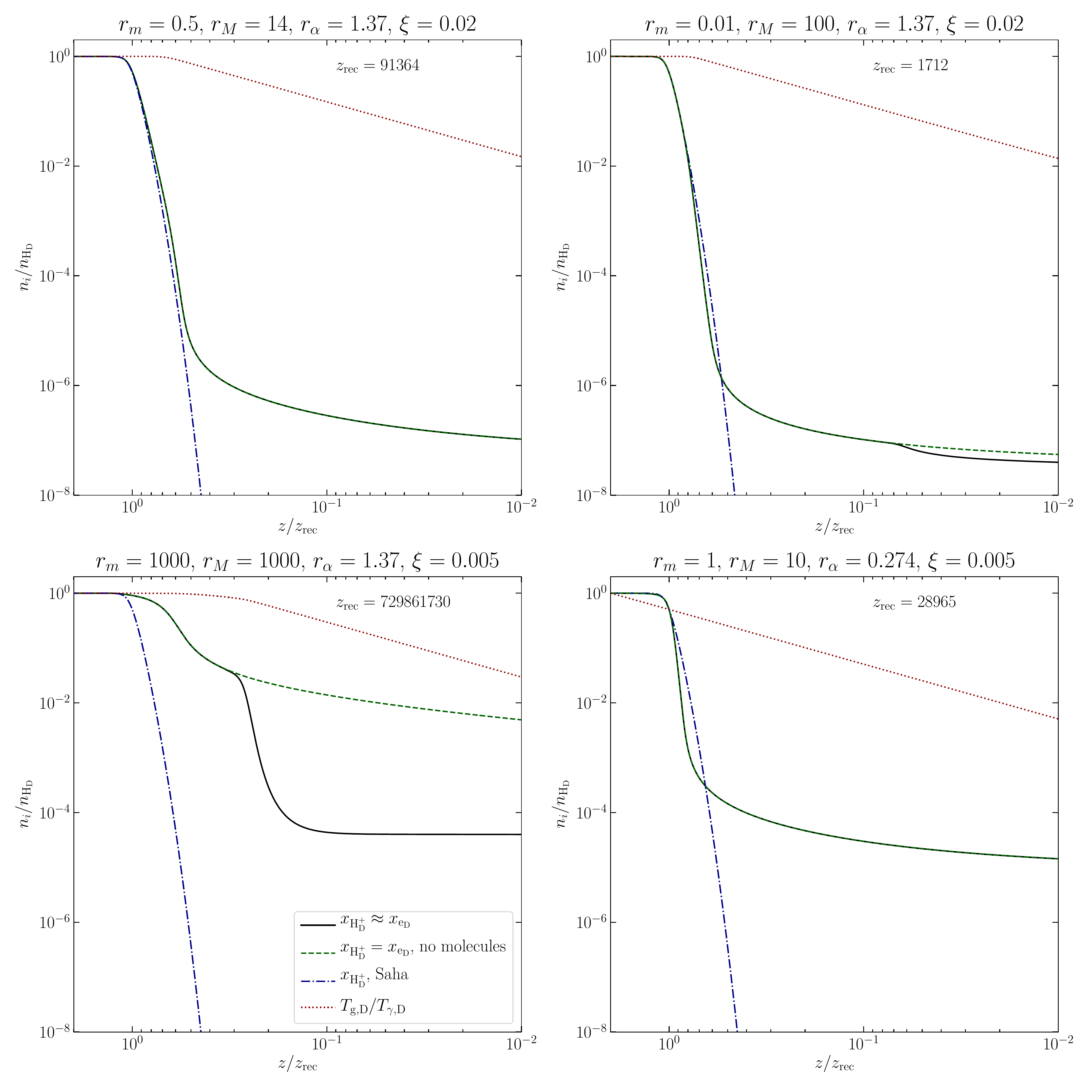}
    \caption{The evolution of the ionization fraction for several choices of parameters: in our full treatment, neglecting molecular states, and in Saha equilibrium. Note that the inclusion of molecules has a quite small effect on the free proton fraction, except in the bottom left. 
    }
    \label{fig:rec_variants}
\end{figure*}

In \reffig{rec_variants}, we show explicitly the evolution of the free ion fraction in our full treatment (black, solid), the 3-level treatment neglecting the formation of molecules (green, dashed), and in Saha equilibrium (blue, dot-dashed) for the same four choices of parameters. Except in the bottom left, the recombination proceeds initially nearly in equilibrium, until the expansion of the universe drives the freeze-out. In the bottom left, the recombination rate is suppressed by the large electron mass ($\alpha_{\rH_\rD} \propto r_m^{-2}r_\alpha^2$), so the recombination rate drops out of equilibrium (i.e.~falls below the Hubble rate) early in the recombination. This leads to early departure from the Saha ionization fraction and a large freeze-out ionization fraction. 

 We also plot $T_{\rm g,D}/T_{\gamma,D}$, illustrating the thermal decoupling of the gas from the dark photons. In the Standard Model, the thermal decoupling is delayed compared to recombination by the low baryon-to-photon ratio, $\eta$. The decoupling in the bottom left is qualitatively most similar to the Standard Model, but the delay is due instead to the large freeze-out free electron fraction and high temperature around recombination. In the bottom right, the weak coupling means that the decoupling precedes recombination, while in the top panels the decoupling occurs during recombination. 

Note that except in the bottom left the molecules have a minimal effect on the ionization fraction at all times. The $\rH_{\rD, 2}$ production pathways do not lead to a net depletion of electrons and (for the parameters considered) they proceed rapidly compared to all other reactions. However, both the formation and destruction of $\rH_{\rD, 3}^+$ lead to a net recombination. A difference is also discernible in the bottom right, due to the mutual neutralization reaction $c_{\rm H7}$ converting some of the $\rH_\rD^-$ and $\rp_\rD$ into neutral hydrogen rather than molecular hydrogen. That reaction rate is enhanced by $r_\alpha^3 r_m^{-3}\approx 10^5$, and still leads to only a factor $\lesssim 2$ difference in the final ionization fraction. We conclude that except when $\hto$ dominates over $x_{\rH_\rD^+}$, the cosmological scales which depend on opacity to Thomson scattering (discussed below) are weakly affected by the inclusion of molecules

\begin{figure*}
    \centering
    \includegraphics[width=\textwidth]{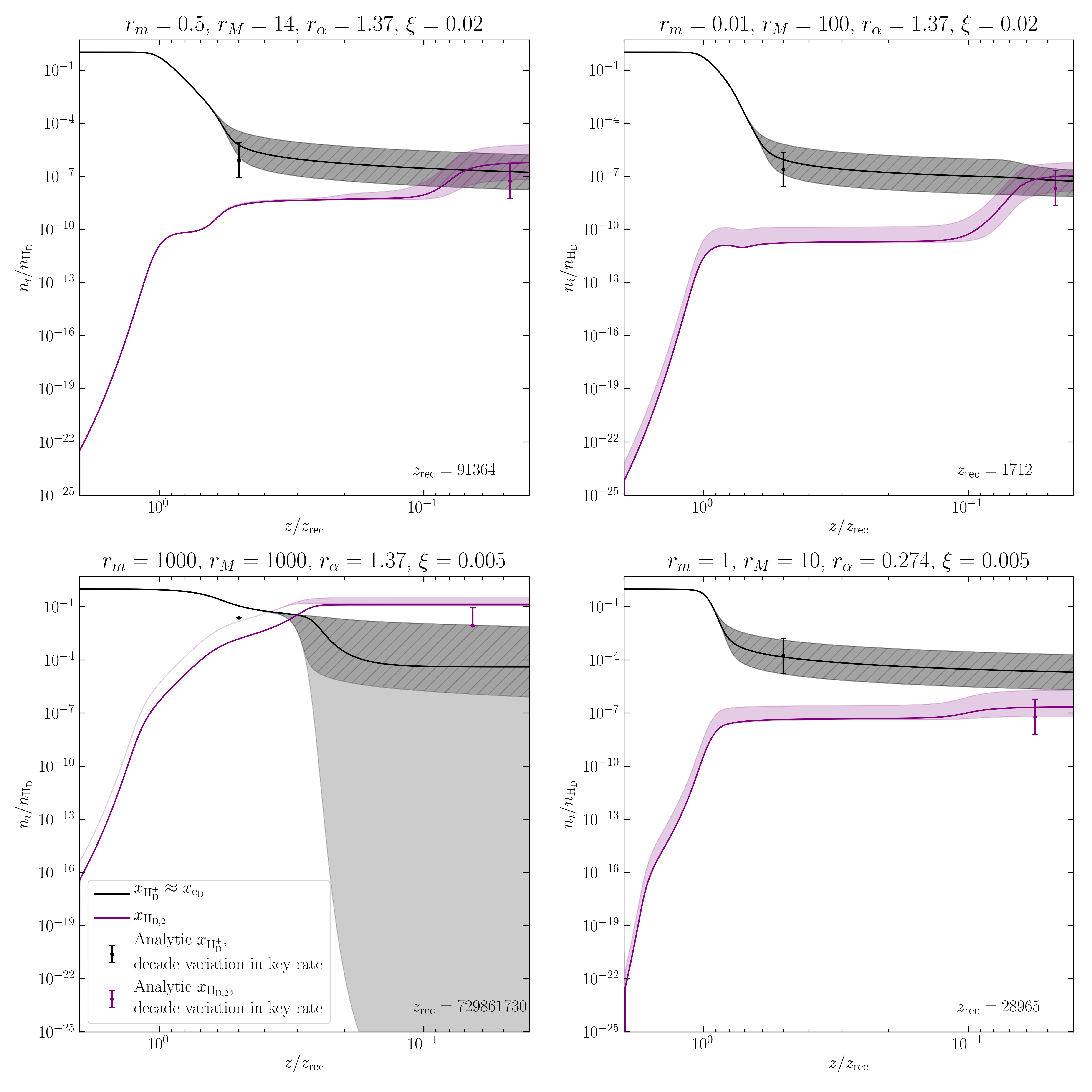}
    \caption{Effects of increasing and decreasing individually the reactions with the largest effect on the final $\hto$ fraction (purple, shaded), $x_{\rp_\rD}$ (gray, shaded), and $x_{\rm e_{\rD}}$ (gray, shaded, hatched), along with the effect inferred from varying those rates by the same amount in our analytic approximation (dots with bars). The results do not sensitively depend on most of the rates involved. See text for discussion of the large difference between $x_{\rp_\rD}$ and $x_{\rm e_{\rD}}$ in the bottom left.}
    \label{fig:uncert}
\end{figure*}

We conclude this section by analyzing the propagation of uncertainties in the rates through our calculation, using our analytic approximation. In Fig.~\ref{fig:uncert}, for each reaction in turn we have artificially first increased and then decreased the rate by a factor of 10. We use the reaction variations which produce the highest and lowest final $x_{\rH_\rD^+}$, $x_{e_\rD}$, and $x_{\rH_{\rD, 2}}$ to plot an uncertainty envelope around each track.  In the same way, by feeding these artificial rates to our analytic model, we can understand how sensitively our results depend on the accuracy of
the reaction rates. 

Except in the bottom left panel, the analytic error bars closely match the numerical error bars. In the bottom left, the estimate for the $\hto$ abundance fails because the high density, $n_{\rH_\rD}$, allows significant molecule formation for a large window around recombination, even in chemical equilibrium. This is not captured by our estimate assuming a short burst of formation with the electron fraction inferred from Saha equillibrium. The estimate for $x_{\rH_\rD^+}$ fails because it does not take into account the $\rH_{\rD, 3}^+$ recombination channel: the reaction which produces the largest effect numerically ($\htp$ formation, $c_{\rH8}$) has no effect in the analytic treatment. We also see a uniquely sensitive dependence of $x_{\rH_\rD^+}$ on this reaction rate. The linear dependence of the freeze-out abundance on the reaction rate revealed in \refeq{CE} holds only when the abundances of the reactants can be treated as constant. Here we see instead the exponential rate dependence of the solution $x_{\rH_\rD^+} \propto \exp[{-t}/{(c_{\rH i}n_{\rH_{\rD}\rm I}})]$. The electron abundance is not so radically affected, because only the lesser of $x_{\rH_\rD^+}$ and  $x_{\rH_{\rD,3}^+}$ is subject to exponential depletion, and once $x_{\rH_{\rD, 3}^+}$ exceeds $x_{\rp_\rD}$, we have $x_{\rm e_\rD} \approx x_{\rH_{\rD, 3}^+}$. 
Fig.~\ref{fig:uncert} tells us that we can be confident in the accuracy of our results to almost exactly the extent we are confident in the rates of these four key reactions. There are no non-linear dependencies where small errors in certain rates leading to large changes in the final results. This mitigates the impact of uncertainties in the Standard-Model reaction coefficients.

\begin{figure}
\includegraphics[width=1.1\linewidth]{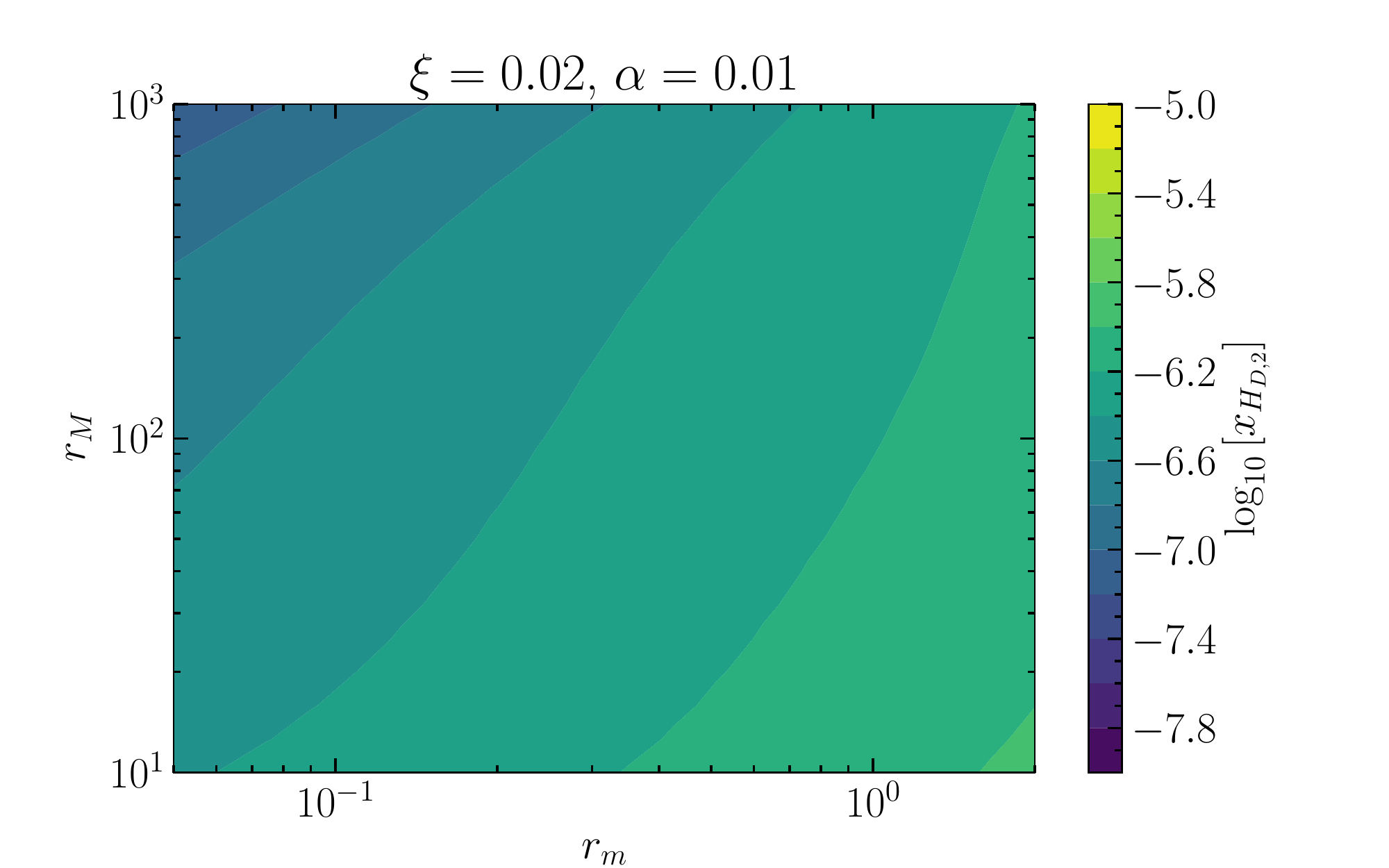}
\includegraphics[width=1.1\linewidth]{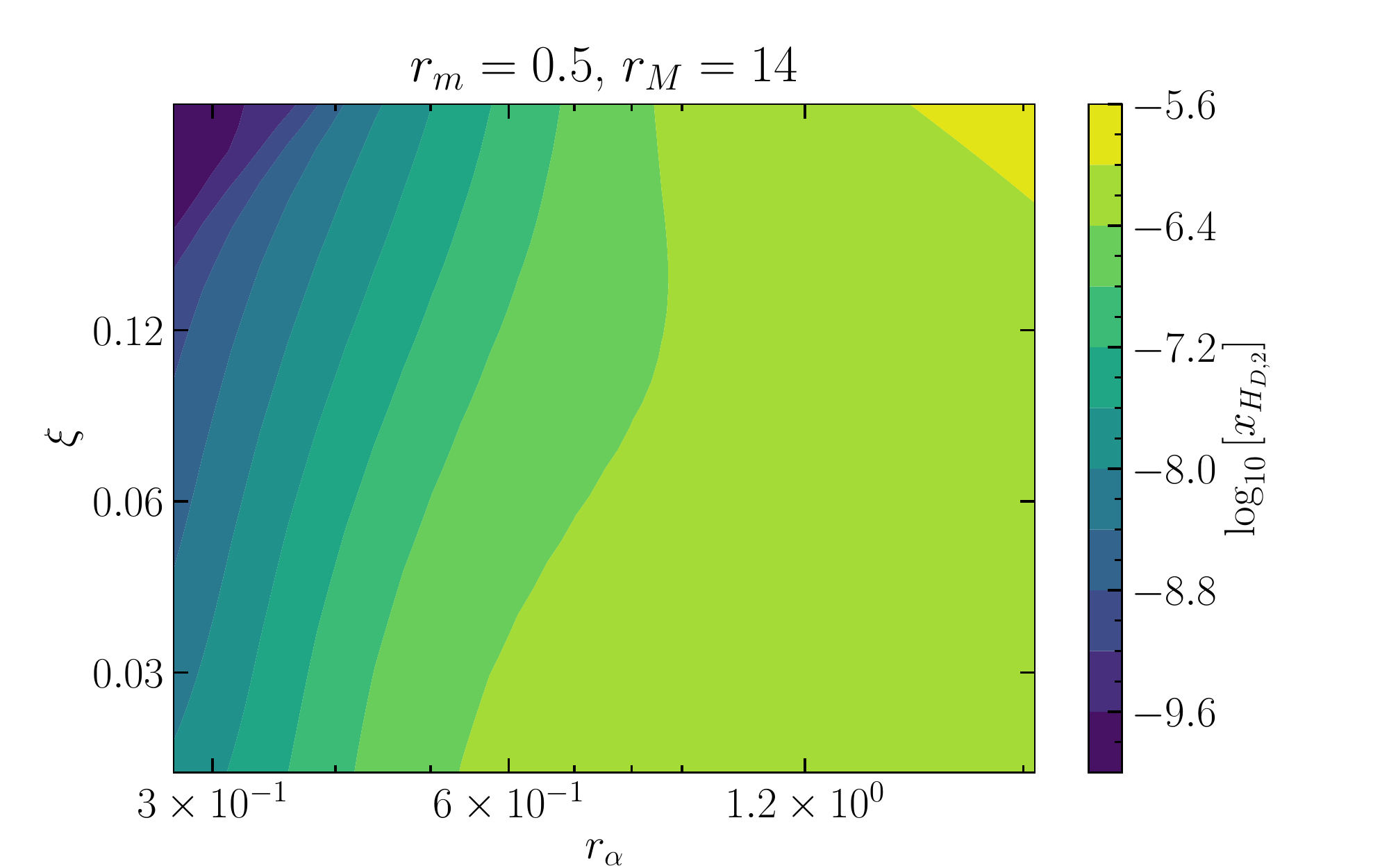}
\caption{The dependence of the molecular hydrogen fraction $x_{\rH_{2, \rD}}$ at redshift 30 on the model parameters. }
    \label{fig:h2frac}
\end{figure}

Now we are ready to show the dependence of the molecular-hydrogen fraction at redshift 30 on the model parameters in Fig.~\ref{fig:h2frac}, by first fixing $\xi$ and $\alpha$ (left panel) while varying the masses, and then fixing the masses while varying $\xi$ and $\alpha$ (right panel). 
 The effect of varying each parameter can be understood through its effect on the number density $n_{\rH_\rD}$ at fixed photon temperature: a higher number density leads to more molecular hydrogen.

\section{Cosmological Consequences}
\label{sec:consequences}
A full exploration of the cosmological consequences of dark matter which can cool is an ambitious program. 
Generating predictions for the merger rate of dark compact objects will require further analytic work and simulations at a wide range of scales. 
In this section, however, armed with a detailed picture of the homogeneous 
evolution of atomic dark matter in the previous sections, we use simple analytic tools to sketch out the large-scale structure in this model.
We shall demonstrate that the model can easily satisfy existing observational constraints and is consistent with related work while estimating the results of future, more detailed investigations. We begin by examining the alterations to the linear matter power spectrum in our model and then proceed to estimate the halo mass function in the Press-Schecter formalism \citep{Press1974}.

\subsection{Linear Power Spectrum: Acoustic Oscillations and Diffusion Damping}
\label{sec:linearpk}
On scales much larger than the horizon size at the dark matter/dark photon decoupling, dissipative dark matter behaves like ordinary cold dark matter. On small scales this is not the case. An exact treatment of the alterations to the linear matter power spectrum in a dissipative dark matter model requires solving the Einstein-Boltzmann equations for the given dark matter model, as in \cite{cyr-racine_cosmology_2013}. Alternatively, one can adopt a phenomenological parameterization of the matter power spectrum and then map the parameters of this model back to the dark matter microphysics, as in \cite{bohr_ethos_2020}. Our approach hews closer to that of \cite{bohr_ethos_2020}, but simpler still.

Atomic dark matter can alter the linear matter power spectrum on small scales primarily through two effects: dark acoustic oscillations (DAO) and dark diffusion (Silk) damping. 
The following two length scales define the pivot length scales below which 
the two effects alter the matter power spectrum.

The DAO scale is given as 
\be
d_{\rm DAO} = 
\int_0^{a_{\rm dec}} c_s(a)\frac{da}{a^2H(a)}
,
\ee
where we calculate the physical sound speed from
\be
c_s(a) = \frac{c}{\sqrt{3(1+R(a))}}.
\ee
as a function of the scale factor $a=1/(1+z)$.
The quantity $R(a)$ is the ratio of the relativistic inertias, $\bar{\rho}+\bar{P}$, 
with the background density $\bar{\rho}$ and pressure $\bar{P}$, of 
the dark matter and dark photon, given as
\be
R(a)
=
\frac{3\bar{\rho}_{DM}}{4\bar{\rho}_{\gamma,D}}
= 
2.32\times 10^{10} a
\(\frac{\Omega_{DM}}{0.25}\)
\(\frac{\xi}{0.02}\)^{-4}.
\,
\ee
Clearly, for a low dark photon temperature ($\xi$), $R(a)$ can be very large. The DAO scale depends only on $R(a)$ (which depends most strongly on $\xi$) and on the scale factor at decoupling, which is controlled dominantly by $E_{\rH_\rD}$ compared to $\xi$.

The diffusion damping scale is given by
\citep{zaldarriaga/harari:1995}
\be
\frac{1}{k_D^2}
=
\int_0^{a_{\rm dec}}
\frac{da}{a^2H(a)}\frac{1}{6(1+R)n_{\re_\rD}\sigma_{\rm T,D}a}
\[
\frac{16}{15} + \frac{R^2}{1+R}
\]
\,.
\ee
Note that even for very small or large values of $R(a)$ the diffusion scale is the geometric mean of the photon mean free path and the horizon size, modified by factors of order unity. A weak dependence on $M$ enters through the number density. 

\begin{figure}
\includegraphics[width=1.1\linewidth]{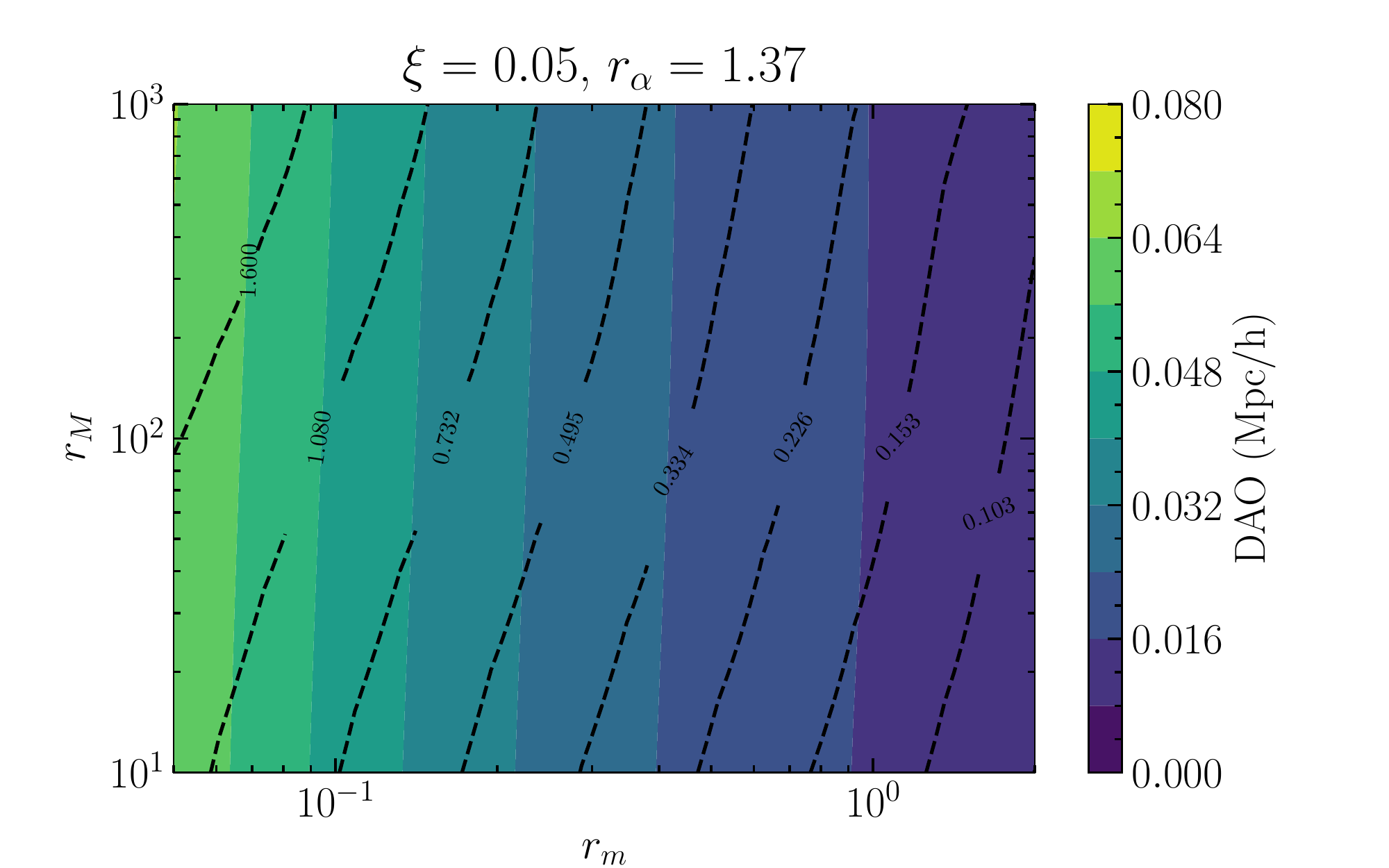}
\includegraphics[width=1.1\linewidth]{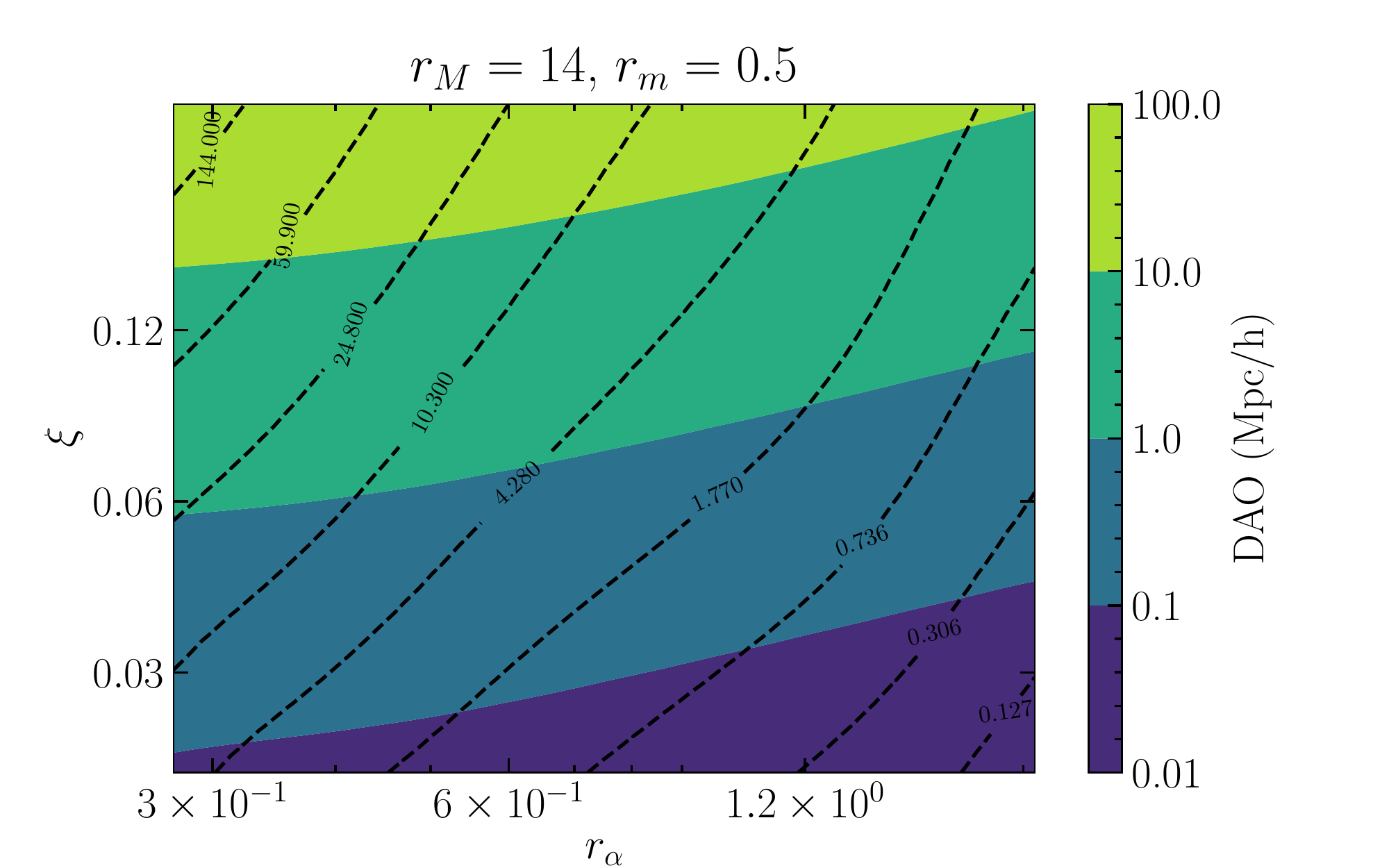}
\caption{The dependence of the diffusion damping scale (black, dashed) and the acoustic scale (color bars) on the model parameters.
}
\label{fig:deps}
\end{figure}

Since the DAO scale is suppressed by $R(a)$ while the diffusion scale is not, for low $\xi$ the diffusion scale can far exceed the DAO scale. In this case, the DAOs are completely washed out by diffusion damping. We show the 
dependence of these two scales on the model parameters in Fig.~\ref{fig:deps}.

Since we find that for the parameters studied in \cite{darkkrome} the diffusion scale is larger than the acoustic scale, we calculate the matter power spectrum assuming CDM using CAMB \citep{Lewis2000} and then apply a Gaussian cutoff (as in e.g.~\cite{Seljak_1996}):
 \begin{equation}
     P(k,z) = P(k,z)_{\rm CAMB}e^{-(k/k_D)^2}.
 \end{equation}
On scales smaller than the diffusion scale, any other modifications to the power spectrum are buried under the exponential diffusion damping. This treatment is sufficient to determine whether the formation of the halos studied in \cite{darkkrome} is suppressed by the diffusion scale, and whether this suppression will appear at observable scales in the matter power spectrum.

\subsection{Implications for Dissipative Dark Halos}
\label{sec:darkhalos}
Equipped with the modified matter power spectrum, we can estimate the halo mass function at various redshifts. In addition to the aforementioned modification to the linear power spectrum, atomic dark matter does not virialize exactly like CDM on small scales. In particular, the dynamics of the collapse will be altered by scattering and subsequent radiative transitions if some of the dark matter is ionized. 

However, early in the nonlinear collapse (which takes approximately a Hubble time), the number density does not differ much from background. Therefore, the ionization fraction will not differ much from the freeze-out value. Since the gas is predominantly neutral, it will behave as CDM through this phase. Just prior to virialization, the number density and temperature rise rapidly, but at this point dissipative effects cannot prevent the formation of the halo or alter its mass. 

However, the eventual shape and structure of these halos may be quite different if they can cool efficiently subsequent to their formation. Hydrodynamical simulations with cooling will be necessary to resolve these effects.

Motivated by this discussion, we adopt the approach which has been developed for warm dark matter (WDM), and has recently been applied to gravity-only simulations of various self interacting dark matter models \citep{bohr2021halo}. Atomic dark matter is similar to WDM in that both models generate a suppression of power at small scales while behaving like CDM on large scales. 

In principle, thermal velocities in WDM models can not only lead to a suppression of power on small scales but can also interfere with the virialization of halos. This has been treated in \cite{Benson_2012} by 
employing a ``moving barrier'', raising the critical density for halos 
smaller than the suppression scale.
Later work \citep{Schneider_2013, Angulo_2013}, however, shows that for simulations initialized in the nonlinear regime the thermal velocity effect can be safely neglected compared to the gravitational velocities. 
That is, the only modification necessary is to the linear matter power spectrum, exactly as in atomic dark matter. 

We begin with the analytic prediction for the halo mass function in Press-Schechter theory \citep{Press1974}. 
There, the cumulative mass function of halos of mass greater than 
$M_{\rm h}$  is given by the fraction of the smoothed linear density field (with 
smoothing radius $R=[3M_{\rm h}/(4\pi\bar\rho_M)]^{1/3}$) exceeding 
the critical value $\delta_c\simeq1.69$ calculated in the spherical collapse
model. Here, $\bar\rho_M$ is the comoving background matter density.
Then, the differential number of halos of mass $M_h$ per volume is given as 
\be
\frac{dn}{dM_{\rm h}} = 
\frac{\bar \rho}{M_{\rm h}^2}\left \rvert \frac{d\ln \sigma_{M_{\rm h}}}{d \ln M_{\rm h}}\right \rvert f(\sigma_{M_{\rm h}}),
\label{eq:halomass}
\ee
with $\sigma_{M_{\rm h}}$ is the root-mean-square of the overdensity smoothed over a 
region with length scale $R$:
\be
\sigma^2_{M_h} =\frac{1}{2\pi^2} \int dk k^2 P(k) W_R^2(k).
\ee
The window function $W_R(k)$ is the Fourier transform of the smoothing 
filter $W_R(r)$, and the Press-Schechter multiplicity function is
\be
f(\sigma_{M_{\rm h}})
=
\sqrt{\frac{2}{\pi}}
\frac{\delta_c}{\sigma_{M_{\rm h}}}
{\rm exp}\left(-\frac{\delta_c^2}{2\sigma_{M_{\rm h}}^2}\right)\,.
\ee
Strictly speaking \citep{bond/etal:1991}, the formalism is only valid for the sharp-$k$ filter: 
\begin{equation}
    W_R(k) = \begin{cases}
        1  & k \leq 1/R\\
        0  & k > 1/R
    \end{cases},
\end{equation}
which has the property that the contributions to $\sigma_{M_h}$ from each successive decrease in $M_h$ are uncorrelated. However, for CDM the results are quite insensitive to the choice of filter, so a real space tophat filter is often employed.
Finally, \cite{Sheth_1999} derived a multiplicity function by fitting to
simulation results, 
\be 
f_{ST}(\sigma)= A \sqrt{\frac{2a}{\pi}}\left[1+\left(\frac{\sigma^2}{a\delta_c^2}\right)^p\right] \frac{\delta_c}{\sigma}\exp{\left(-\frac{a\delta_c^2}{2\sigma^2}\right)},
\ee
with $p= 0.3$, $A = 0.322$, $a = 0.707$ \citep{stiavelli},
which has been widely used in the literature.

The problem with applying the Press-Schechter formalism to atomic dark matter is that,
for models with a cut-off in the power spectrum, the halo mass function 
becomes sensitive to the choice of the smoothing filter. 
\cite{Schneider_2013} found that the sharp-$k$ filter 
(which is anyway the only theoretically justified choice) produces the best
agreement with simulations when modifying the mass-to-smoothing-radius
relation as
\be
M_{sk} = \frac{4\pi}{3}\bar\rho[c R_{sk}]^3,
\ee
where $c=2.7$ from calibration to their simulations, and setting $a=1$ for the $a$ parameter in the Sheth-Tormen multiplicity function.
Note that warm dark matter simulations including \cite{Schneider_2013} have been plagued by force resolution issues which have led to spurious small-scale structures. This was largely resolved in \cite{Angulo_2013}, which justifies using the Press-Schechter formalism with modifications described by \cite{Schneider_2013} to obtain a qualitatively correct halo mass function. However, that work cautions that further effort is needed to obtain a halo mass function which is quantitatively accurate for scales near and below the cutoff in the matter power spectrum. These caveats apply equally to atomic dark matter. 

\begin{figure}
    \includegraphics[width=1.05\linewidth]{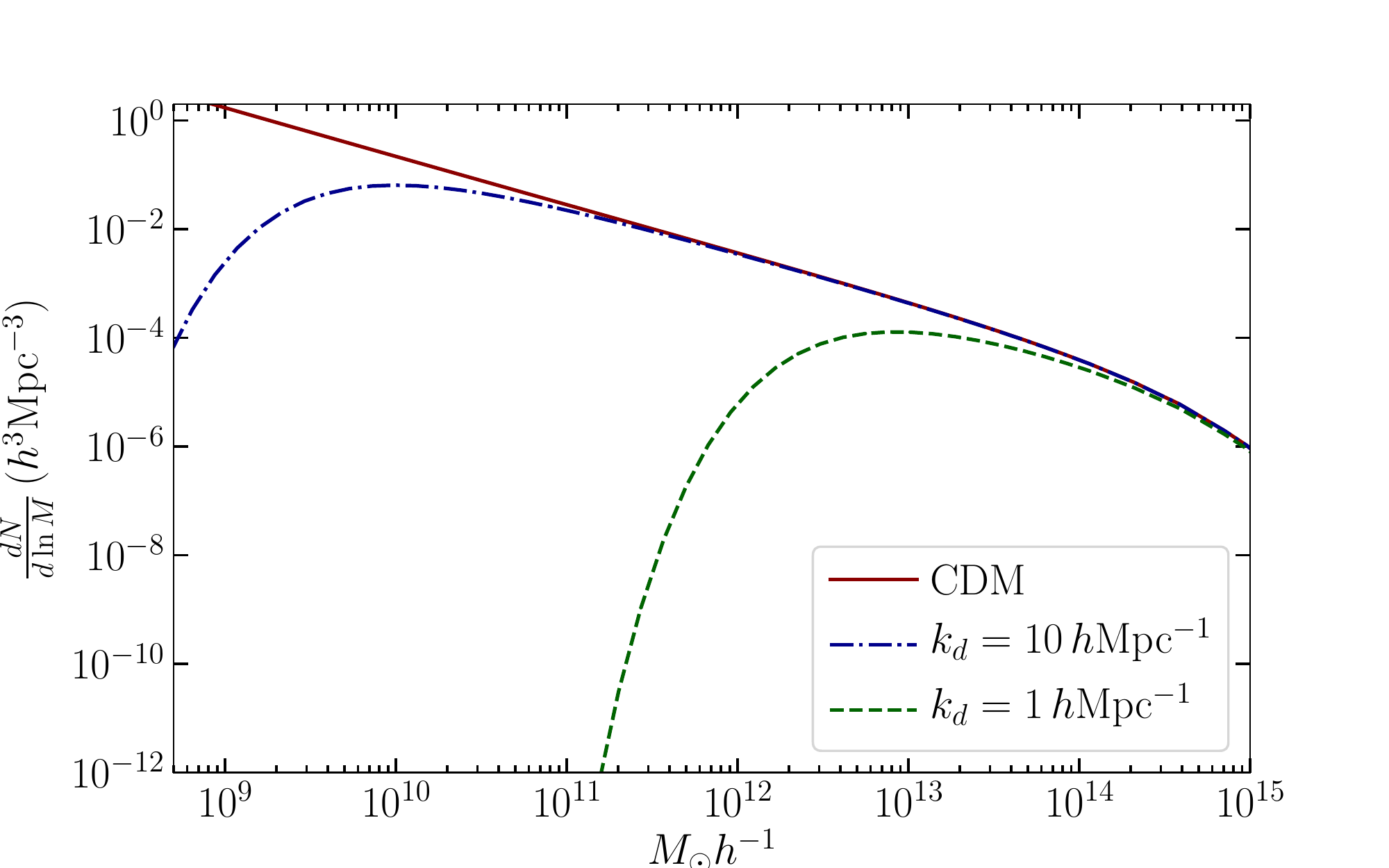}
    \includegraphics[width=1.05\linewidth]{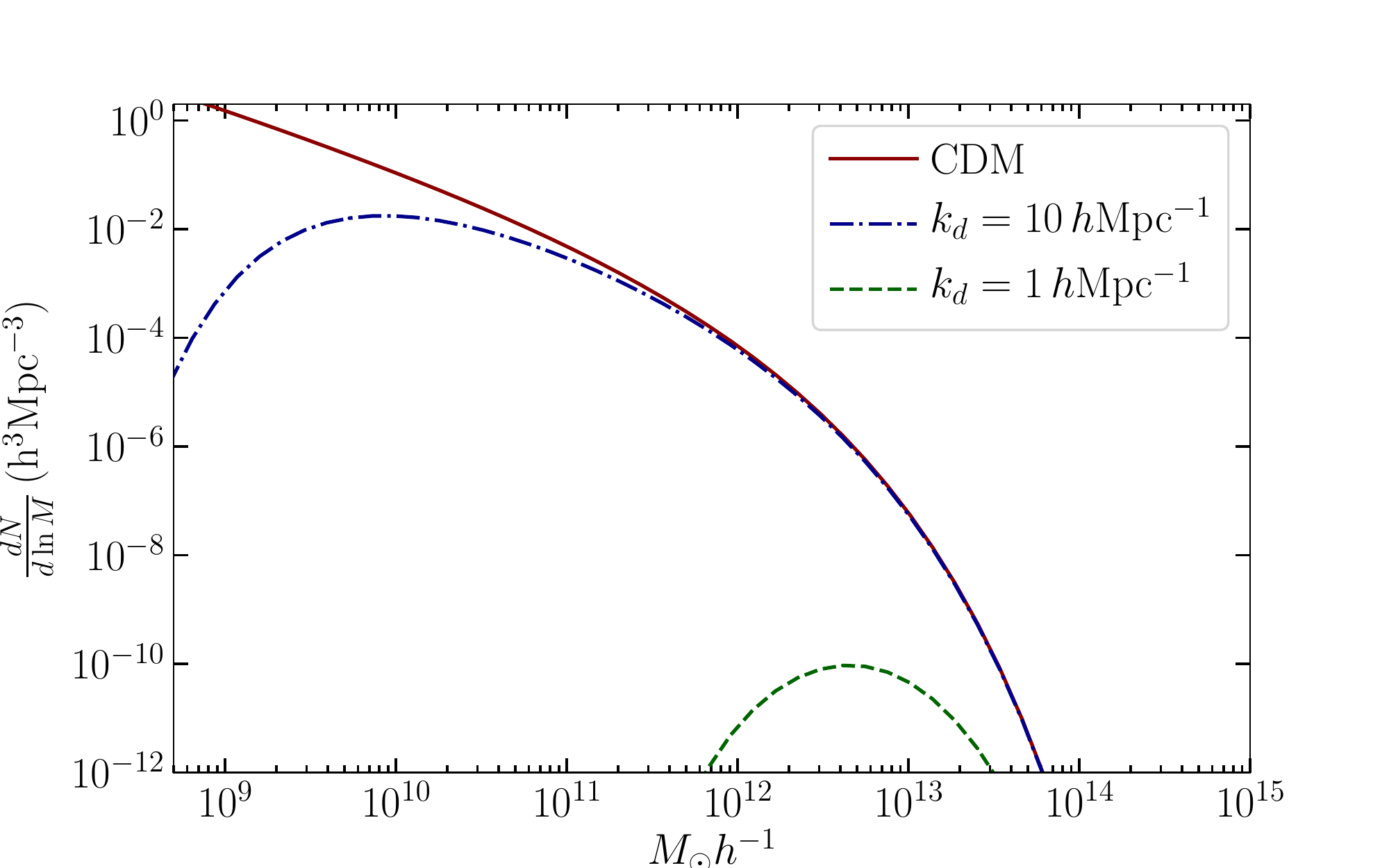}
\caption{The halo mass function at $z= 0$ (upper) and $z=5$ (lower). The Gaussian cutoff in the power spectrum leads to a corresponding suppression in the halo mass function at small scales. Clearly, $k_d =1$ is ruled out by the existence of $\sim 10^{11} \, \rm M_\odot$ galaxies. We include it here to illustrate the dependence of the halo mass function on $k_d$.}
\label{fig:hmf}
\end{figure}

In Fig.~\ref{fig:hmf}, we show the halo mass function $z=0$ (top panel) and $z=5$ (bottom panel) for the usual CDM case and for the atomic dark matter case with 
$k_d=10{h/{\rm Mpc}}$ (blue, dot-dashed line)
and $k_d=1{h/{\rm Mpc}}$ (green, dashed line).
The diffusion scale is the only relevant variable affecting the halo mass 
function.

\section{Conclusion}
\label{sec:conclusion}

In this paper, we have calculated the background evolution of the abundance of 
dark molecules and ions in the atomic-dark-matter model.
To do so, we use the {\rm Recfast++} software and the interaction-rate 
coefficients re-scaled by the methods derived in \citet{darkchem}. Previous work in the mirror matter model (equivalent to atomic dark matter with $\ra{}=\rc{}=\rx{}=1$ for our purposes) \citep{latif_black_2019, damico_massive_2018} assumed the ratio $x_{e} /x_{H_{2}}\approx 10^{2}$ carried over from the Standard Model, independent of the dark matter parameters. Here, we calculate the primordial molecular abundance directly and find that this assumption often fails, because the primordial molecular abundance depends not only on the primordial ionization fraction but also on the dark-particle number density at the time of molecule formation.  

We have also presented an analytic calculation for the final 
dark-molecular abundance estimate, which agrees well with the full numerical
result. The analytic estimate also indicates that the final abundance is at worst linearly sensitive to the uncertainties in the reaction rates. 

We find that when dark recombination happens at high redshift, 
large dark particle number density promotes the molecular processes so that most of the atomic dark matter particles ends up in molecules. However, as long as the molecule formation is fast compared to the recombination (which is true for all cases studied here) the molecular processes hardly change the recombination history, thereby, the DAO scale and the dark diffusion (Silk) damping scale.

We have also estimated the linear power spectrum of atomic dark matter by calculating the DAO scale and dark-diffusion (Silk) damping scale. When the dark photon temperature is much lower than the CMB temperature, the dark diffusion scale is much larger than the DAO scale, so that we neglect the DAO effect and modify the linear power spectrum by a Gaussian damping factor due to diffusion.

Finally, using the modified linear matter power spectrum, we have estimated the halo mass function using the Press-Schechter formalism modified to fit the warm dark matter simulation results.
This work provides the initial conditions and some useful estimates for 
the results of full atomic dark matter simulations with chemistry, 
beginning with the companion paper, \cite{darkkrome}.

In the future, hydrodynamical simulations with radiative cooling will provide insight into the implications of atomic dark matter at scales ranging from the largest structures in the universe to the smallest compact objects. Although precise results for such complex nonlinear physics will require significant investment, this program will begin to promote a broad class of alternative dark matter models\textemdash all those in which the dark matter forms hydrogen-like bound states\textemdash to testable theories whose parameters can 
be constrained by observation.

\acknowledgements
The authors thank Jens Chluba for providing a special version of the
{\sf Recfast++} code and for useful discussions about the calculation of 
cosmic recombination history and spectral distortions. We thank the anonymous referee for providing a truly excellent and professional report, which contributed significantly to the depth of analysis presented in the revised version.
Funding for this work was provided by the Charles E. Kaufman Foundation of the Pittsburgh Foundation.
\software{Recfast++ \citep{chluba_recombinations_2010}, Julia \citep{bezanson2012julia},  DifferentialEquations library \citep{DifferentialEquations.jl-2017}}

\appendix
\section{Atomic dark matter implementation of {\sf Recfast++}}\label{sec:app}

The {\sf Recfast++} code calculates the temperature and ionization history of Standard Model hydrogen and helium. In order to implement the evolution of atomic dark matter states using the {\sf Recfast++}, besides changing atomic rate coefficients as described in \refsec{implement}, we have re-defined the input variables as follows.
With these redefinitions, the same {\sf Recfast++} code can be applied to both the standard recombination and the atomic dark matter scenario.
\begin{itemize}
{\item
We have preserved the basic {\sf Recfast++} functionality, including helium recombination, but set the helium fraction to zero.}
{\item
The {\sf Recfast++} variable representing the CMB temperature $T_0$ is reinterpreted as the dark photon temperature today, parameterized by $T_{\gamma, D} = \xi T_0$.}
{\item
The number density of the dark matter is calculated as 
\begin{equation}
    n_{H_D}(z) = \frac{3H_0^2\Omega_{DM}}{8 \pi G M }(1+z)^3\,,
    \label{eq:nhd}
\end{equation}
where $M$ is the mass of dark matter proton.
In {\sf Recfast++}, the hydrogen number density is parameterized by $\Omega_b$. To match Eq.~(\ref{eq:nhd}), we set
\begin{equation}
    \Omega_b \rightarrow \Omega_{DM} r_M^{-1}\,,
\end{equation}
where $r_M = M/m_p$.
This is equivalent to setting the input {\it dark-matter} density parameter 
\be
\Omega_{DM} \to \Omega_{b} + (1-r_M^{-1})\Omega_{DM}\,,
\ee
because we need to keep the total matter density
$\Omega_m = \Omega_{DM}+\Omega_b$ equal to the usual $\Lambda$CDM cosmology value. Since {\sf Recfast++} takes $\Omega_m$ as an input, the modification
to $\Omega_{DM}$ is only implicit.
}
{\item
Finally, we have included all the relativistic degrees of freedom other than the dark photons into $N_{\rm eff}$:
\be
N_{\rm eff} \rightarrow \frac{1}{\xi^4}\left[N_{\rm eff} + \frac87 \left(\frac{T_{\rm CMB}}{T_\nu}\right)^4\right]\,,
\ee
so that the total radiation energy density becomes
\begin{align}
\rho_{\rm rad}
        =
        \rho_{\gamma} + \rho_{\nu} + \rho_{\rD,\gamma}
        =&
\frac{\pi^2}{15}
\left[
    1 
    + 
    \frac78N_{\rm eff}\left(\frac{T_\nu}{T_{\rm CMB}}\right)^4
    +
    \left(\frac{T_{\gamma,\rD}}{T_{\rm CMB}}\right)^4
\right] T_{\rm CMB}^4
\vs
        =&
\frac{\pi^2}{15}
\left[
    1 
    + 
    \frac78
    \frac{1}{\xi^4}
        \left(
        N_{\rm eff}
        +\frac87
        \left(\frac{T_{\rm CMB}}{T_{\nu}}\right)^4
        \right)
        \left(\frac{T_\nu}{T_{\rm CMB}}\right)^4
\right] T_{\gamma,\rD}^4\,.
\end{align}
}
\end{itemize}

\bibliographystyle{aasjournal}
\bibliography{darkrec.bib}

\end{document}